\documentclass[twocolumn,showpacs,
preprintnumbers,
amsmath,
amssymb,
amsfonts,
aps,
prb,
superscriptaddress
]{revtex4}
\usepackage{amssymb}
\usepackage{amsmath}
\usepackage[pdftex]{color}
\usepackage[pdftex]{graphicx}
\usepackage{bm}
\usepackage{fancyhdr}
\usepackage[utf8]{inputenc}
\usepackage[francais]{babel}

 \newcommand\ST{\rule[1.0em]{0pt}{0.5em}}
\newcommand{\bc}{\begin{center}}
\newcommand{\ec}{\end{center}}

\def\R{\mathcal{R}}
\def\S{\mathbf{S}}

\def\Tr{\textrm{Tr}}
\def\n{\mathbf{n}}

\def\eps{\epsilon}
\def\<{ < \mathrm{k}_F}
\def\>{ > \mathrm{k}_F}
\def\eps{\epsilon}

\newcommand{\ud}{\textrm{d}}

\newcommand{\Matthree}[9]{ 
\left( \begin{array}{ccc}  
#1& #2 & #3  \\ 
#4 & #5 & #6  \\
#7 & #8 & #9 \\
 \end{array} \right) }
 \newcommand{\Vectthree}[3] { \left( \begin{array}{c}  
#1\\ 
\ST
#2  \\ 
 \ST
#3  \\
 \end{array} \right)}

\newcommand{\ZZ}{$\mathbb{Z}_2 \times \mathbb{Z}_2$}

\begin{document}
\title{Quantum phase transitions in three-leg spin tubes}
\author{D. Charrier}
\affiliation{Max-Planck-Institute f\"ur Physik komplexer Systeme, N\"othnitzer Strasse 38, D-01187 Dresden, Germany}
\affiliation{Laboratoire de Physique Th\'eorique, Universit\'e de Toulouse, UPS, (IRSAMC), F-31062 Toulouse, France}
\affiliation{CNRS, LPT (IRSAMC), F-31062 Toulouse, France}
\author{S. Capponi}
\affiliation{Laboratoire de Physique Th\'eorique, Universit\'e de Toulouse, UPS, (IRSAMC), F-31062 Toulouse, France}
\affiliation{CNRS, LPT (IRSAMC), F-31062 Toulouse, France}
\author{M. Oshikawa}
\affiliation{Institute for Solid State Physics, University of Tokyo, Kashiwa 227-8581, Japan}
\author{P. Pujol}
\affiliation{Laboratoire de Physique Th\'eorique, Universit\'e de Toulouse, UPS, (IRSAMC), F-31062 Toulouse, France}
\affiliation{CNRS, LPT (IRSAMC), F-31062 Toulouse, France}

\date{\today} 
\begin{abstract}
 We investigate the properties of a three-leg quantum spin tube using
 several techniques such as the density matrix renormalization group
 method (DMRG), strong coupling approaches and the non linear sigma
 model (NL$\sigma$M).
 For integer $S$, the model proves to exhibit a particularly
 rich phase diagram consisting of an ensemble of $2S$ phase
 transitions. They can be accurately identified by the behavior
 of a  non local string order parameter associated to the breaking of a
 hidden symmetry in the Hamiltonian. The nature of these
  transitions are further elucidated within the different approaches.
  We carry a detailed DMRG analysis in the specific cases $S = 1$. The
  numerical data confirm the existence of two Haldane phases with
  broken hidden symmetry separated by a trivial singlet state. The
  study of the gap and of the von Neumann entropy suggest a first 
  order phase transition but at the close
  proximity of a tricritical point separating a gapless and a first
  order transition line in the phase diagram of the quantum spin tube.
\end{abstract}
\maketitle
%
%

\section{\label{sec:Introduction} Introduction}
Frustrated spin models in one dimension have attracted attention for
both the uniqueness of their characteristics and the diversity of
their properties. In contrast to higher dimensional spin systems,
quantum spin chains have no long range order. If there is no
frustration, the properties of the chain are essentially governed by
the parity of the spin: the Heisenberg spin chain for instance has a
gapless spectrum and algebraic correlations when the value of the spin
is a half-integer whereas it has a gap and exponentially decaying
correlations when the spin is an integer~\cite{Haldane1983}. When
frustration is present, the problem gets much more complex and the
possibilities for the low-energy spectrum are also broadened. An
illustrative example is given by the spin ladder with $S = 1/2$ and
additional diagonal couplings. Depending on the strength of the
frustrating couplings, the ground state of the system can be described
in terms of rung singlets, short-range valence
bonds~\cite{Topoladders}, or would eventually
dimerize~\cite{Starykh2004}. The transitions between some of these
phases have been proposed to be deconfined quantum critical points
which could support fractionalized spinons~\cite{Kim2008}.

Another family of problems concerns the integer spin ladders. The
comprehension of integer spin chains have considerably improved since
the discovery
of the AKLT Hamiltonians~\cite{AKLT} and 
the early work of den Nijs and Rommelse~\cite{Nijs}.
In particular, the ground state
of the spin-$1$ Heisenberg chain is now well understood: it displays a
subtle hidden topological
degeneracy~\cite{Kennedy,Kenn/Tasaki,Tasaki}, associated to a
non-vanishing non-local string order parameter~\cite{Nijs} and
supports edge states. The question of the preservation of the
topological order when couplings between different chains are
introduced is an open issue. It is believed that this order should be
highly sensitive to perturbations.
As a matter of
fact, a simple coupling between two spin-$1$ chains leads rapidly to
the destruction of the topological order~\cite{Todo},
reflecting the fragility of the edge states towards the perturbation (see also Ref.~\onlinecite{Anfuso}.) 
However, it is
also possible to maintain the topological phase by adding frustrating
nearest-neighbor interactions. In this case, a direct first-order
transition between two different topological phases can be
observed~\cite{Kolezhuk}. The question of the stability of the
topological order in spin ladders is of crucial importance if one
thinks of these systems as intermediates between $1d$ and $2d$ systems
and regards them as a pathway to discover a spin liquid behavior in
two-dimensional systems.

In this work, we investigate the presence and nature of topological
phases in an asymmetric three-leg quantum spin tube with
integer spin quantum numbers. 
The triangular spin tube has already been extensively
studied in the spin-$1/2$ case. Abelian bosonisation
techniques~\cite{Schultz} arguments suggest that the system is gapped
when the tube is symmetric and maximally frustrated.
It is interesting to introduce asymmetry
among the coupling in each triangle. 
The model with the asymmetry has been studied by 
density-matrix-renormalization-group (DMRG) algorithm. Recent DMRG
calculations~\cite{Sakai} have demonstrated that the dimer order is
unstable against a small but non zero anisotropy coupling, 
that eventually drives the system into a critical phase.

Much less is known
in the case of integer spins. The triangular geometry provides a
simple and natural way to introduce frustration, and we thus hope to
find unconventional behaviors. Here, one coupling between two legs
is varied, thus controlling the strength of the frustration
(see Figure \ref{fig:tube1}) in order to explore
a large phase diagram.
The possibility of quantum phase transitions with deconfined
spinons is also an interesting question.

Besides DMRG, a group of methods that can be used to investigate this
problem are the large-$S$ approaches. Among them is the non linear
sigma model (NL$\sigma$M) which furnishes crucial information
regarding the spectrum of spin chains and ladders~\cite{Haldane1983,
Sierra}. Spin models with triangular geometry are described in the
continuum by a $SO(3)$ rotation matrix field, in contrast to collinear
antiferromagnets for which the NL$\sigma$M theory involves a single
unit vector field~\cite{Delduc}. $SO(3)$ NL$\sigma$M are characterized
by the absence of a topological term in the action~\cite{Dombre} and,
in $d > 1$, by a non-trivial fixed point with an enlarged $SO(4)$
symmetry~\cite{Azaria}. Even without topological term, integer and
half-integer spin behave differently due to the occurrence of
topological defects~\cite{Rao,Haldane1988}. It remains to be seen how
this scheme is perturbed by the introduction of an anisotropy in the
triangular geometry.

We determine the phase diagram of the anisotropic spin tube with
integer spin $S$ by gathering together the results obtained from
diverse methods: strong coupling expansion, large-$S$ approaches and
DMRG. We find that the tube supports $2S$ quantum phase transitions
when the anisotropic coupling is varied. The nature of the transitions
is debated. We begin in section \ref{sec: Modelandlimit} 
with the proper definition of the model and
introduce its strong coupling limit. Different phases are delimitated
depending on the value of the quantum spin $J$ of each triangle. In
section \ref{sec: IntegerS}, we develop the notion of string order
parameter and we show how the spin tube model can be rewritten in
terms of a local Hamiltonian with a discrete $\mathbb{Z}_2 \times
\mathbb{Z}_2$ symmetry. This hidden symmetry is broken when $J$ is odd
and remains unbroken when $J$ is even. To understand the nature of the
phase transition, we turn in the third part to the large-$S$
approaches. We begin with a spin-wave analysis to determine the
low-energy modes of the model. We then derive the
NL$\sigma$M and the associated Renormalization Group (RG)
equations. In our derivation, we put a careful emphasis on the
evaluation of the total Berry phase of the tube. We find $2S$ special
values of the anisotropic coupling corresponding to a non trivial
Berry phase. Then, we focus on the special case of the spin-$1$ tube
with a strong coupling approach and a DMRG study. The DMRG results
reveal the presence of two quantum phase transition points,
in adequacy with the
predictions of the strong coupling limit. The order of the transition
is proposed to be first order but the numerical data also strongly
suggest the proximity of the system to a tricritical point. Finally, we
provide a numerical phase diagram for the spin-$2$ tube where various
even/odd $J$ phases compete.

\section{\label{sec: Modelandlimit}
The model and some simple limits}

\subsection{\label{subsec: Model}
The model}

\begin{figure}
 \begin{center}
\includegraphics*[width=7cm]{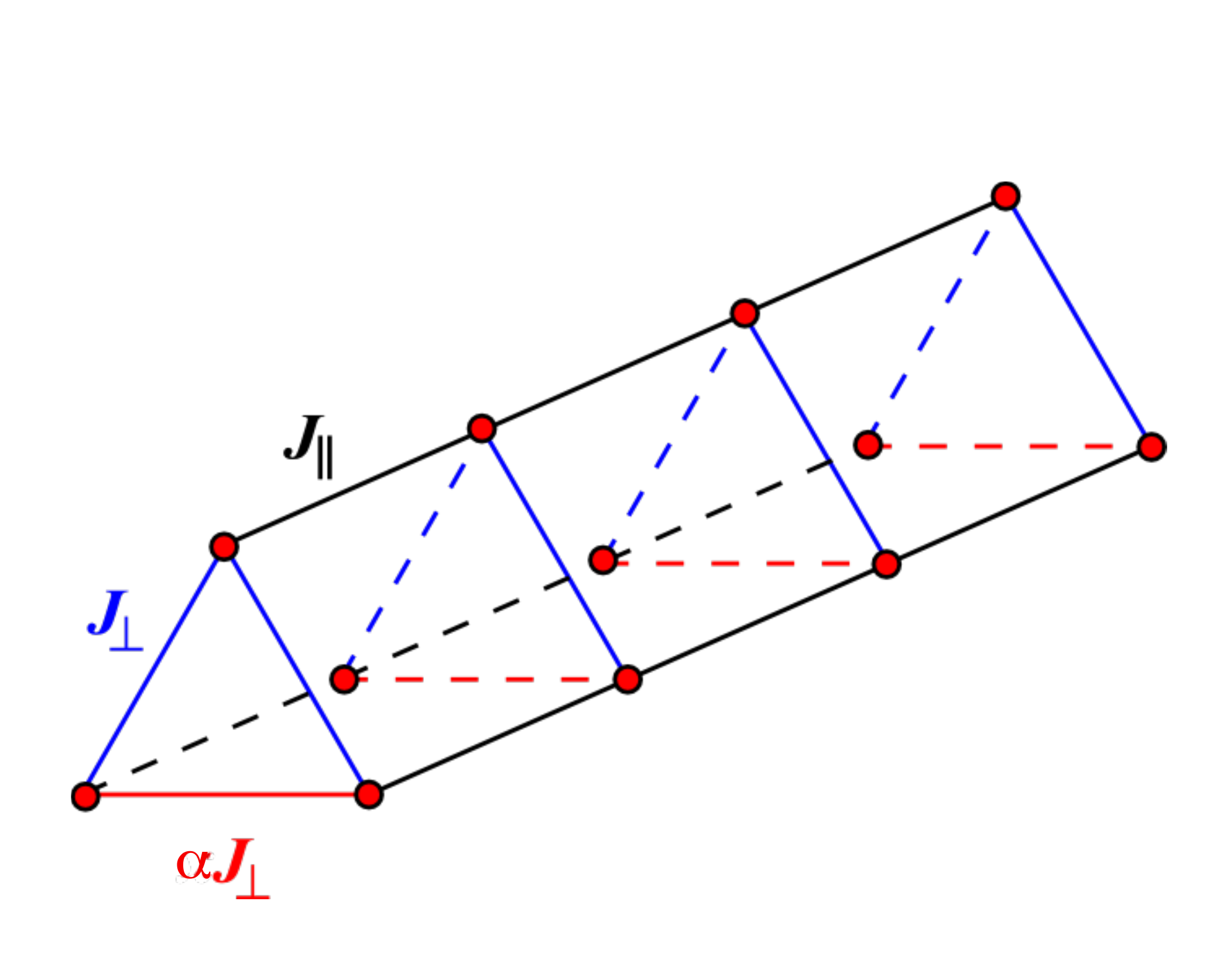}
\caption{(color online) The three different parameters defining the coupling of the three-leg spin tube.}
\label{fig:tube1}
\end{center}
\end{figure}

The anisotropic triangular spin tube is a quantum ladder problem defined by three relevant parameters  (Figure \ref{fig:tube1}): the parallel coupling $J_{\parallel}$, the perpendicular coupling $J_{\perp}$ and the anisotropy parameter $ 0 \leq \alpha $. The Hamiltonian reads:
\begin{eqnarray}
\hat{H} &=&  \hat{H}_{\parallel}+\hat{H}_{\perp} \label{eq:Hamiltonian} \\
 \hat{H}_{\parallel} &=& J_{\parallel}\sum_{i,a} \left( \mathbf{S}_{i,a} \cdot  \mathbf{S}_{i+1,a} \right) \nonumber \\
\hat{H}_{\perp} &=& J_{\perp} \sum_i \left( \mathbf{S}_{i,3} \cdot \mathbf{S}_{i,1} +  \mathbf{S}_{i,2} \cdot \mathbf{S}_{i,3} + \alpha \mathbf{S}_{i,1} \cdot \mathbf{S}_{i,2} \right), \nonumber
\end{eqnarray}
with $i = 1,\ldots ,N$ being the intra-chain index and $a = 1,2,3$ being the rung index. 
The point $\alpha =0$ corresponds to the unfrustrated open ladder while $\alpha = 1$ is also special because of its translation symmetry in the transverse direction.

\subsection{\label{subsec: Decoupledclas}
The classical case}

We start by determining the classical configurations of spins which minimize the energy of each triangle by replacing the spin operators $\S_a$ with classical vectors $S\n_a$. 

\begin{figure}
 \begin{center}
\includegraphics*[width=6cm]{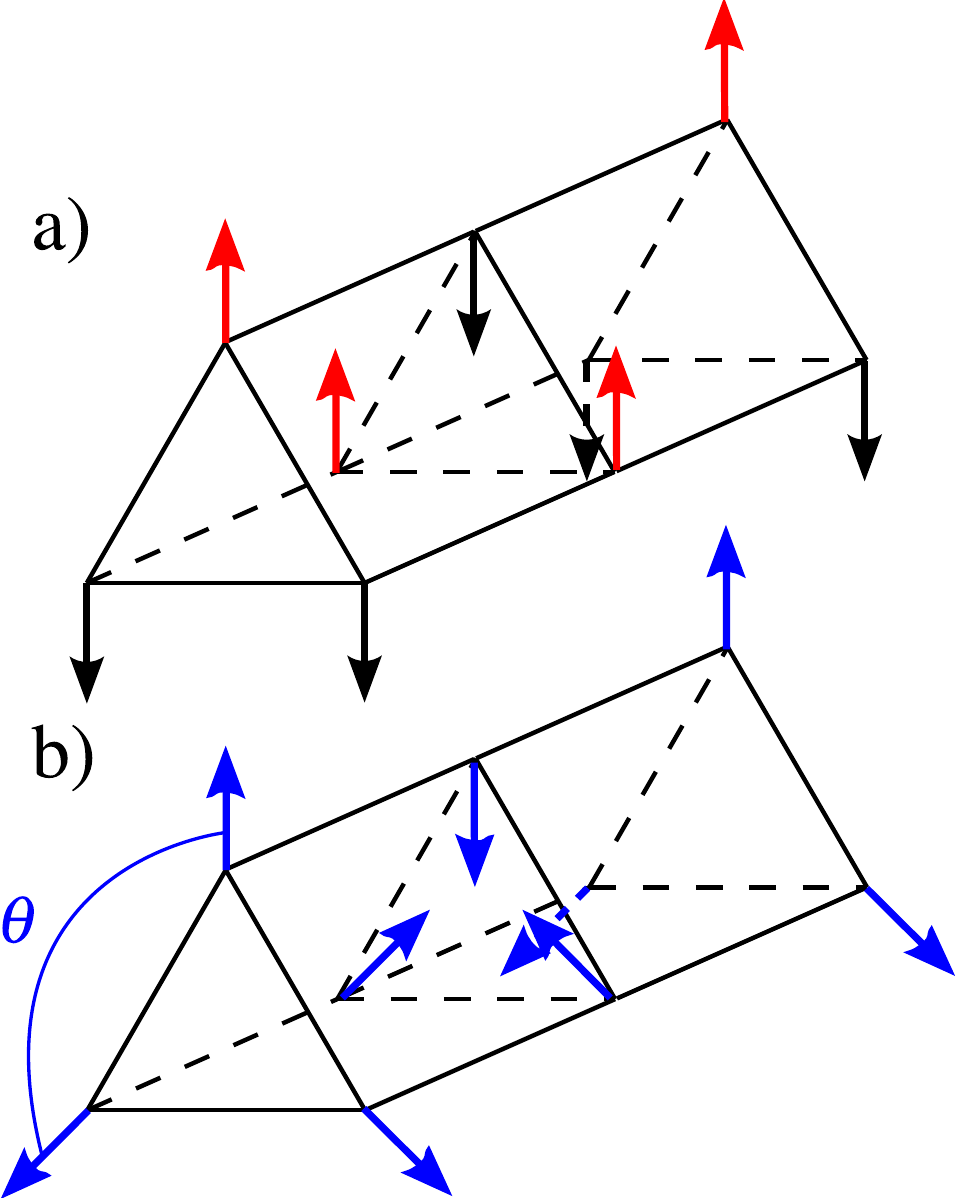}
\caption{(color online) Top: The collinear configuration which minimizes the energy for $0 \leq \alpha \leq 0.5$. Bottom: the coplanar configuration which minimizes the energy for $0.5 \leq \alpha$. }
\label{fig:tube2}
\end{center}
\end{figure}

For $\alpha \geq 0.5$ the solutions that minimize the energy are of the
kind of the coplanar solution of Figure~\ref{fig:tube2} (b).
\begin{equation}
\n_{1} =\Vectthree{\sin \theta}{0}{\cos \theta},  \n_2 =   \Vectthree{-\sin \theta}{0}{\cos \theta},  \n_{3} = \Vectthree{0}{0}{1},
\label{eq:coplanar}
\end{equation}
with:
\begin{equation}
\cos \theta  = -\frac{1}{2\alpha}.
\end{equation}
In the extreme limit $\alpha \rightarrow \infty$, the two vectors
$\n_1$ and $\n_2$ point in opposite direction and the third spin is
essentially free. The system reduces then to the problem of one single
chain. On the opposite, decreasing $\alpha$ one enters the regime $0
\leq \alpha \leq 0.5$ in which the lowest energy state is an alternated
collinear configuration of Figure~\ref{fig:tube2}(a).
In this regime the physics becomes the one of an open unfrustrated ladder.

Note that the collinear state and (\ref{eq:coplanar}) are both
continuously degenerate but have a different degree of degeneracy. For
$0 < \alpha \leq 0.5$, any alternated collinear configuration minimizes
the energy. Thus, choosing a ground state is equivalent to picking up an
oriented axis. For $0.5 < \alpha$, all the classical ground states
are given by a global rotation of the triad $(\n_1,\n_2,\n_3)$.
This, in turn, requires to choose an oriented axis and an angle.

\subsection{\label{subsec: Decoupledquant}
Quantum spins : the decoupled limit}

\begin{figure}[!h]
\begin{center}
\includegraphics*[width=7cm]{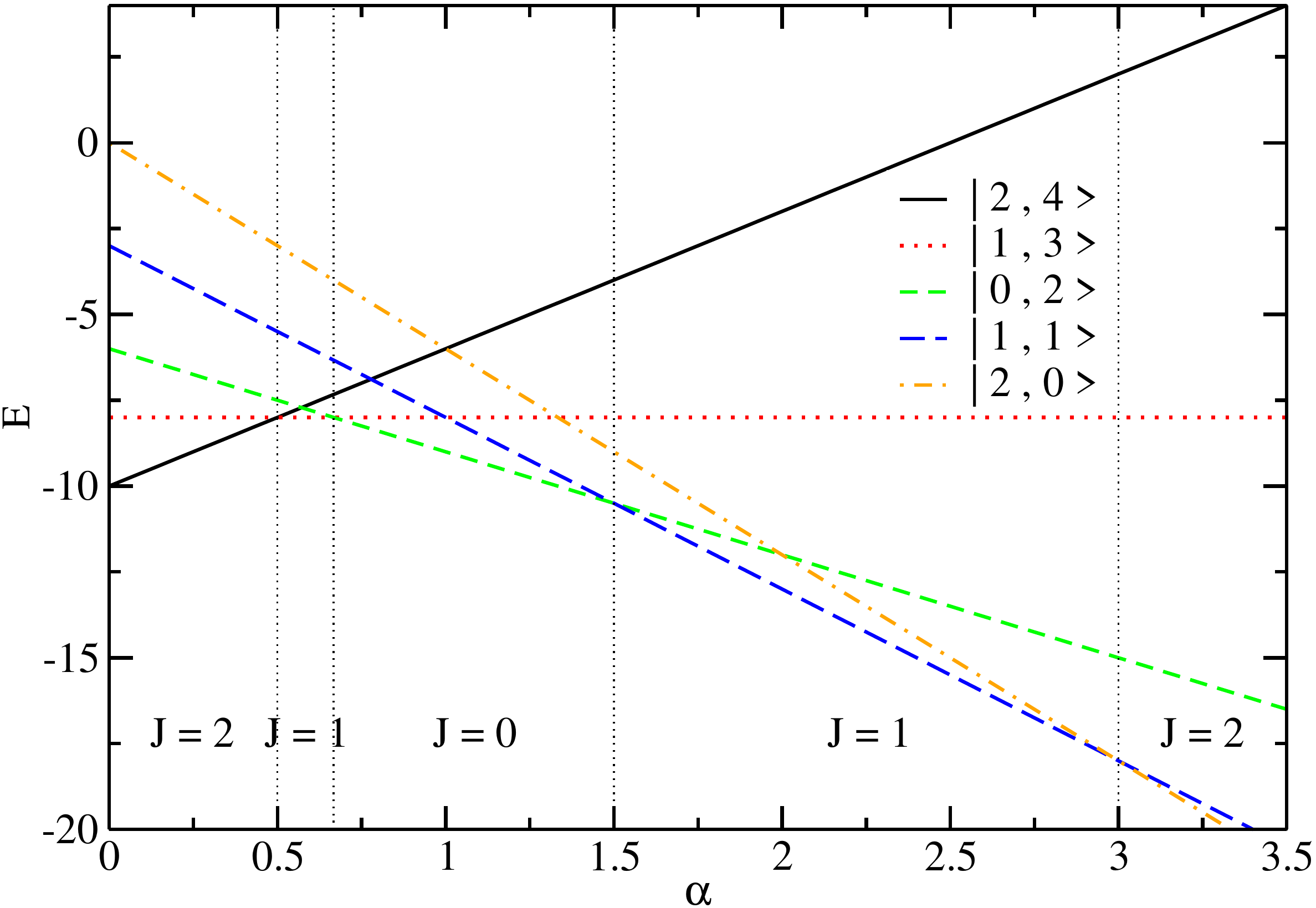}
\caption{(color online) Spectrum of a single triangle for $S = 2$ as a function of $\alpha$ as given by the two sequences (\ref{eq:sequence1}) and (\ref{eq:sequence2}). The higher energy levels are not represented.}
\label{fig:EL}
\end{center}
\end{figure}

Introducing the triangle spin $\mathbf{J} = \S_1 + \S_2 + \S_3$ and the bond spin $\S_{12} = \S_1 + \S_2$, the rung Hamiltonian reads:
\begin{eqnarray*}
&& \mathbf{S}_{i,3} \cdot \mathbf{S}_{i,1} +  \mathbf{S}_{i,2} \cdot
 \mathbf{S}_{i,3} + \alpha \mathbf{S}_{i,1} \cdot \mathbf{S}_{i,2}
 \nonumber \\ &=& \frac{J^2}{2} +(\alpha -1) \frac{{S_{12}}^2}{2}
 -(2\alpha + 1)\frac{S(S+1)}{2}\\
 &=& \frac{J(J+1)}{2} + (\alpha -1) \frac{S_{12}(S_{12}+1)}{2} -(2\alpha + 1)\frac{S(S+1)}{2}
\end{eqnarray*}
where we have replaced the spin operators by their eigenvalues. To
determine the ground state, we need to label each state by their value
of total spin $J$ and their intermediate spin $S_{12}$. For $\alpha =
1$, the $S_{12}$ levels are degenerate and the ground state is
obtained for the smallest value of $J$. Thus, the ground state is the
singlet state $| 0, S \rangle$ (if $J = 0$, $S_{12} = S$
necessarily). When turning on the anisotropy, other levels will
compete with this state. It is straightforward to show that the
sequence of ground states $| J, S_{12} \rangle$ between $\alpha =1$
and $\alpha = 0$ is:
\begin{equation}
| 0, S \rangle \rightarrow | 1, S +1 \rangle \rightarrow ... \rightarrow | S-1, 2S-1 \rangle \rightarrow | S , 2S \rangle
\label{eq:sequence1}
\end{equation}
The first level crossing happens for $\alpha = \frac{S}{1+S}$. The
last level crossing occurs at $\alpha = 0.5$. From this last result we
can conclude that both, classically and quantum mechanically, the
point $\alpha = 0.5$ corresponds to the entrance into the unfrustrated
open ladder regime given by $\alpha = 0$. On the other side of the
isotropic point, $\alpha \geq 1$, 
the sequence of ground state is given by:
\begin{equation}
|0,S \rangle \rightarrow |1,S-1 \rangle \rightarrow ... \rightarrow | S-1, 1 \rangle \rightarrow |S,0 \rangle
\label{eq:sequence2}
\end{equation}
The first crossing takes place at $\alpha = \frac{1+S}{S}$ and the
last one occurs at $ \alpha = 1+S$. After this point, the triangle
consists of two spins coupled into a singlet and an isolated spin.
For instance,
for $S = 1$, there is a level crossing at $\alpha = 0.5$ between the
singlet state $| 0 ,1 \rangle$ and the triplet $| 1, 2 \rangle$ and
another one at $\alpha = 2$ between the singlet and the triplet
$|1,0\rangle$. On Figure \ref{fig:EL}, we plot the evolution of the
main energy levels for $S = 2$.

Thus, in the strong rung coupling limit $J_{\parallel}=0$,
there are $2S$ transition points in $0 \leq \alpha \leq \infty$.
If we now add a small longitudinal coupling $J_{\parallel} \ll J_{\perp}$, 
we can still expect that $2S+1$ different phases are
present. However, we need to know how to effectively make the
distinction between them. As we will see in the next chapter, this can
be achieved with a non local string order parameter. The nature of the
phases corresponding to $J = 0$ and $J \neq 0$ are clearly different. In
the former case, the tube just consists of a trivial superposition of
singlets. We will refer to this phase as the singlet phase. In the
latter, the properties of the tube are more similar to those of a single
chain with $S = J$ spins, and we will refer to them as Haldane-like
phases.

\section{\label{sec: IntegerS}
Integer $S$ case and hidden symmetry}

\subsection{\label{subsec: Stringorder}
Hidden symmetry and string order parameters}

In order to characterize the different phases 
suggested by the strong coupling analysis, 
we would like to find a suitable order
parameter enabling us to describe the phase transitions.
Usually, different phases are characterized by (local) order parameters,
which detect spontaneous symmetry breakings.
However, in some cases this standard approach does not work.
This includes, in particular, the Haldane phase of $S=1$ chain:
it has no local order parameter but still is a distinct phase
separated from a trivial phase by a quantum phase transition.
In order to characterize the Haldane phase,
the non-local ``string order parameters''~\cite{Nijs}, one of which is
\begin{align}
\lim_{|k-i| \rightarrow \infty}
\left \langle
S^z_i \exp \left( i \pi \sum_{l = i}^{k-1} S_l^z \right) S^z_k
\right \rangle ,
\label{eq:string_chain}
\end{align}
is useful.
It has been confirmed that it is non-vanishing within the
Haldane phase but is zero in a trivial phase
(for example the large anisotropy phase of Ref.~\onlinecite{Tasaki}).

The problem with the non-local order parameter such as
Eq.~\eqref{eq:string_chain}, in general, is that
it is not quite clear if there is necessarily a phase
transition between two states, when a non-local order parameter
vanishes in one state but is non-zero in the other.
Kennedy and Tasaki~\cite{Kenn/Tasaki} clarified the meaning of the
string order parameter~\eqref{eq:string_chain}, as
an order parameter measuring a spontaneous breaking
of {\em hidden} discrete symmetry.
Namely, there exists a non-local unitary transformation
which transforms the Hamiltonian $\cal H$
to a Hamiltonian $\tilde{\cal H}$ with short-range
interaction and with a discrete \ZZ\ symmetry.
The \ZZ\ symmetry is hidden in a non-local way in the
original Hamiltonian $\cal H$.

The string order parameter~\eqref{eq:string_chain}
is transformed by the same non-local unitary
transformation to the standard ferromagnetic order parameter.
Thus, non-vanishing string order parameter~\eqref{eq:string_chain}
implies a spontaneous breaking of the hidden \ZZ\ symmetry.
The spontaneous symmetry breaking clearly distinguishes the
phases.
In this sense, the string order parameter indeed qualifies
as an order parameter, despite its nonlocality.

The hidden \ZZ\ symmetry breaking also implies
4-fold groundstate degeneracy.
This appears contradictory to the uniqueness of the groundstate
in the Haldane phase.
However, the non-local unitary transformation only works
for the open boundary condition.
Thus the hidden \ZZ\ symmetry breaking implies
4-fold groundstate degeneracy of the original Hamiltonian
$\cal H$ \emph{only in the open boundary condition}.
This degeneracy actually corresponds to the existence
of the edge states with spin $S_b=1/2$ at both ends.

The appearance of the edge states can be
understood~\cite{AKLT,Kennedy,Nijs,Tasaki, Kenn/Tasaki}
in the Valence Bond Solid (VBS) picture where
each spin $1$ is seen as a triplet of spins $1/2$.
Spin $1$ at each site is first decomposed into two spin $1/2$'s.
Each constituent spin $1/2$ is then coupled to
a spin $1/2$ in the neighboring site to form a singlet.
This would give a simple dimerized state of a spin $1/2$ chain.
However, projection to the triplet sector within each site
gives a nontrival state for $S=1$ chain.
If we consider this state on a finite chain with the open
boundary condition, unpaired spin $1/2$ is left free at
each end. Namely, spin $1/2$ degree of freedom appears at
the ends.
The above construction actually gives exact groundstates
for a special, solvable Hamiltonian.
For other models, the constructed state is of course not
an exact groundstate. However, the appearance of the edge
states is a common feature within the $S=1$ Haldane phase.
As a consequence of the edge states,
the groundstates of an open chain is asymptotically 4-fold
degenerate. As mentioned above, this corresponds to the
spontaneous breaking of the hidden \ZZ\ symmetry.

Thus, the hidden \ZZ\ symmetry breaking characterizes
the $S=1$ Haldane phase, unifying the string order parameter
and the edge states.
However, it should be also noted that
this picture is only valid in the presence of the global
\ZZ\ symmetry.
We will, in Sec.~\ref{sec: Conclusion}, discuss from the
perspective of recent, more general characterization of
the Haldane phase\cite{GuWen,Pollmann09a}.

Now let us move on to our problem of the spin tube with
integer spin.
Naturally, ladders/tubes are more complicated than
the single chain, and
various generalizations of the string
order parameter have been proposed.
However, as we have discussed for the single chain,
generally there is no guarantee that
a non-local ``order parameter'' really qualifies as
an order parameter.
Therefore, in this paper, we first
generalize the hidden \ZZ\ symmetry to the tube.
Then we identify the corresponding string order parameters,
which detect spontaneous breaking of the hidden \ZZ\ symmetry.

Following Ref.~\onlinecite{Oshikawa}, the Kennedy-Tasaki transformation
generalized to the tube can be written as:
\begin{equation}
V = \prod_{j < k} \exp \left( i \pi J^z_j J^x_k \right)
\label{eq:transform}
\end{equation}
with $\mathbf{J}_i = \S_{i,1} + \S_{i,2} +\S_{i,3}$.
We impose the open boundary condition on the tube (along the
leg direction).

It is straightforward to show that the spin operators transform into:
\begin{align*}
V S^x_{i,a} V^{-1} &=  S^x_{i,a}  \prod_{i < k}\exp \left( i \pi J^x_k \right) \\
V S^y_{i,a} V^{-1} &= \prod_{k < i} \exp \left( i \pi J^z_k \right) S^y_{i,a}  \prod_{i < k}\exp \left( i \pi J^x_k \right) \\
V S^z_{i,a} V^{-1} &= \prod_{k < i} \exp \left (i \pi J^z_k \right) S^z_{i,a}
\end{align*}

The natural generalization that comes to mind is to define the
two string order parameters:
\begin{align} 
\langle \mathcal{O}^x \rangle
&= \lim_{|k-i| \rightarrow \infty} \left \langle J^x_i \exp \left( i \pi \sum_{l = i+1}^{k} J_l^x \right) J^x_k \right \rangle 
\label{eq:string-order1} \\
\langle \mathcal{O}^z \rangle
&= \lim_{|k-i| \rightarrow \infty} \left \langle J^z_i \exp \left( i \pi \sum_{l = i}^{k-1} J_l^z \right) J^z_k \right \rangle  
\label{eq:string-order2}      
\end{align}
Applying the unitary transformation (\ref{eq:transform}), they reduce to
the local ferromagnetic order parameters:
\begin{equation}
\langle \tilde{\mathcal{O}}^a \rangle
= \langle V \mathcal{O}^a V^{-1} \rangle
= \lim_{|k-i| \rightarrow \infty}
 \langle J^a_i J^a_k \rangle,
\end{equation}
for $a=x,z$.
Now, let us consider the Hamiltonian. This transforms into:
\begin{align}
\tilde{H} &= \tilde{H}_\parallel +  \tilde{H}_\perp \\
\tilde{H}_\parallel &=  J_{\parallel}  \sum_i \left( \sum_{a = 1,2,3} S^x_{i,a} S^x_{i+1,a} \right) \exp( i \pi J^x_i) \nonumber \\
&+ \left( \sum_{a = 1,2,3} S^z_{i,a} S^z_{i+1,a} \right) \exp( i \pi J^z_{i+1}) \nonumber \\
&+ \left( \sum_{a = 1,2,3} S^y_{i,a} S^y_{i+1,a} \right) \exp( i \pi (J^x_i + J^z_{i+1})) \nonumber \\ 
\tilde{H}_\perp &= J_\perp \left( \S_{i,1} \cdot \S_{i,3} + \S_{i,2} \cdot \S_{i,3} + \alpha \S_{i,1} \cdot \S_{i,2} \right) \nonumber 
\label{eq:TransfHam}
\end{align} 
Note that the rung part is invariant under the non-local
transformation. The new Hamiltonian still consists of local
interactions but the global continuous $SU(2)$ symmetry of the
original Hamiltonian has been hidden and only a discrete
$\mathbb{Z}_2 \times \mathbb{Z}_2$ symmetry remains explicit:
it is now only
invariant under the rotation of all spins around the $x$ and $z$ axis
by an angle of $\pi$.

We suggest that this ``hidden'' (non-local) symmetry and
its associate string order parameters delineate the phases of the
system. In the strong coupling limit ($J_{\parallel} \ll J_{\perp}$),
the phase diagram of the spin tube consists of $2S+1$
phases, labelled by the spin index $J$, analogous to the Haldane state
for a spin-$J$ chain. It has been demonstrated by one of
us~\cite{Oshikawa} that not all Heisenberg spin chains, but only the ones
with $J$ odd, do break the hidden $\mathbb{Z}_2 \times
\mathbb{Z}_2$ symmetry and possess a non zero string order
parameter. Thus, as the anisotropy parameter $\alpha$ is varied in the
spin tube (with $J_\perp \gg J_\parallel$),
we will encounter a succession of phases with the
string-order parameters
(\ref{eq:string-order1})-(\ref{eq:string-order2}) alternatively
vanishing and non vanishing.

It is also interesting to consider a disorder parameter
which detects {\em unbroken} hidden \ZZ\ symmetry, given as
\begin{equation}
\langle \mathcal{O}_D \rangle
\equiv \lim_{|i-j| \rightarrow \infty}
\left\langle
\exp{\left( i \pi \sum_{l=i}^{j-1} J^z_l \right)}
\right\rangle .
\label{eq:disorder}
\end{equation}
In fact, this was introduced in Ref.~\onlinecite{RiseFall} as
a ``parity correlation function'' and shown to vanish
in the Haldane phase but non-vanishing
in a trivial phase.
Here we discuss Eq.~\eqref{eq:disorder} from a different viewpoint
from that in Ref.~\onlinecite{RiseFall}.

The non-local transformation~\eqref{eq:transform} maps
the non-local disorder parameter Eq.~\eqref{eq:disorder} to
itself:
\begin{align}
\langle V \mathcal{O}_D V^{-1} \rangle & =
\lim_{|i-j| \rightarrow \infty}
\left\langle \exp{\left( i \pi \sum_{l=i}^{j-1}
e^{i \pi \sum_{k<l} J^z_k}
\tilde{J}^z_l \right)} \right\rangle \notag \\
&= 
\lim_{|i-j| \rightarrow \infty}
\left\langle \exp{\left( i \pi \sum_{l=i}^{j-1} J^z_l \right)}
\right\rangle ,
\end{align}
where we used the fact that $\exp{(\pm i \pi J^z)} = \exp{(i \pi J^z)}$
because $J^z$ only takes integer values.

This correlation function can be interpreted as follows.
The global $\pi$-rotation of spins (in the transformed basis)
about $z$ axis is a generator of the \ZZ\ symmetry.
Let us consider a localized operation, namely
$\pi$-rotation about $z$ axis only
on the spins in the finite section between $i$ and $j$.
This ``localized symmetry generator'' is no
longer a symmetry generator of the system.
We apply this operation to the groundstate,
and the overlap with the groundstate is measured.
The limit $|i-j| \rightarrow \infty$ is taken afterwards.
If the \ZZ\ symmetry is spontaneously broken, the application of
the ``localized symmetry generator'' flips the order parameter
in the finite section. Thus the overlap with the groundstate
asymptotically vanishes in the limit $|i-j| \rightarrow \infty$.
Therefore, the disorder parameter~\eqref{eq:disorder}
vanishes if the hidden \ZZ\ symmetry is spontaneously broken.
This is quite analogous to the well-known disorder parameter
in the quantum transverse Ising chain.~\cite{Kogut}
On the other hand, it does not vanish generically in a trivial phase
where the hidden \ZZ\ symmetry is unbroken.

The discussion here implies that Eq.~\eqref{eq:disorder}
acts as a disorder parameter for the hidden \ZZ\ symmetry,
when the hidden \ZZ\ symmetry is well-defined.
It would be the case even if the inversion (parity) symmetry
is explicitly broken in the Hamiltonian, when the original
argument in Ref.~\onlinecite{RiseFall} does not apply.
(Although here we discussed the case of the tube,
the same argument about the disorder operator
applies to integer spin chains.)

\subsection{\label{subsec: Edgestates} Edge states}

Possible quantum phases of the spin tube may be characterized
by the hidden \ZZ\ symmetry breaking (or non-breaking).
As in the case of single chain,
spontaneous breaking of the hidden \ZZ\ symmetry
implies 4-fold degeneracy of the groundstates,
but only in the open boundary conditions.
This implies the existence of the edge state
(with half-integer spin, if the edge spin quantum number
is well-defined.)

It also implies that, we can investigate whether the
hidden \ZZ\ symmetry is spontaneously broken or not,
by analyzing the existence of the edge states.
If there are no edge states, the hidden \ZZ\ symmetry
cannot be spontaneously broken.
Existence of the edge state would suggest spontaneous
breaking of the hidden \ZZ\ symmetry.
However, it should be noted that the edge states could
appear by a different mechanism unrelated to the
hidden \ZZ\ symmetry.

The existence of the edge states can be analyzed easily
in the strong-coupling limit ($J_{\parallel} \ll J_{\perp}$).
In the strong coupling limit, we can project
to the groundstates of each rung, which changes
according to the sequence~\eqref{eq:sequence1},
as the tube anisotropy parameter $\alpha$ is varied.

Let us first discuss the $S=1$ tube.
In the isotropic
regime ($J_{\parallel} \ll J_{\perp}$ and $\alpha \sim 1$), each
triangle tends to form singlets, and we thus expect a unique
ground-state (with no boundary degeneracy) corresponding to the phase
with unbroken $\mathbb{Z}_2 \times \mathbb{Z}_2$. As this phase has no
spin at the boundary, it will be referred as the $S_b=0$ phase.
On the other hand, in the anisotropic, ``unfrustrated'' regime
($J_{\parallel} \ll J_{\perp}$ and $\alpha \lesssim 0.5$),
the three spins of each
triangle couple to form a $J = 1$ spin object. 
The resulting physics
is essentially that of the spin-1 chain and we expect a ground-state
degeneracy due to the boundary spins. In the language of the
transformed Hamiltonian this corresponds to the broken $\mathbb{Z}_2
\times \mathbb{Z}_2$ phase with $\langle \mathcal{O}^\alpha \rangle
\neq 0$.
We will call this phase the $S_b=1/2$ phase in relation with the spin
$1/2$ edge state of a spin-$1$ chain.

We can also discuss the edge states in 
the weak coupling limit ($J_{\parallel} \gg J_{\perp}$) with an
heuristic argument.
Taken separately, each chain of the tube is
gapped, thus we can assume that a weak interchain coupling will not
change qualitatively the physics in the bulk. On the contrary,
solitary edge excitations of each of the three chains are expected to
be very sensitive to any perturbation. As soon as a coupling $J_\perp$
is introduced, they will be bounded, leaving a unique spin-$1/2$
degree of freedom at each edge. This again corresponds to the
$S_b=1/2$ phase, suggesting broken $\mathbb{Z}_2 \times \mathbb{Z}_2$.
This picture is valid for basically any non-zero value of
$\alpha$. The only special point is $\alpha=1$ where the translation
symmetry in the transverse direction leads to a bigger ${1
  \over 2} + {1 \over 2}$ degeneracy space as it happens with three
spin-$1/2$ with identical AF couplings.

\subsection{Hidden \ZZ\ symmetry breaking in the weak coupling limit}
\label{sec:hiddenZZ_weak}

The edge state analysis implies that the hidden
\ZZ\ symmetry is spontaneously broken in the $S=1$ tube,
in the weak coupling limit $J_\perp \ll J_\parallel$ for
any value of $\alpha$.

However, there is a subtle issue in the weak coupling limit.
It can be shown that the string order
parameters~\eqref{eq:string-order1} and \eqref{eq:string-order2}
exactly vanish at the decoupling point $J_\perp = 0$.
At this point, the groundstate is given by a simple
product of the groundstates of each chain.
The string order parameter~\eqref{eq:string-order1}
can be decomposed as
\begin{align}
 \langle \mathcal{O}^z \rangle &= \lim_{|k-i| \rightarrow \infty}
\left\langle S^z_{i,1}
\exp{\left( i \pi \sum_{l = i}^{k-1} S_{l,1}^z \right)}
S^z_{k,1} \right\rangle_1  \times \notag \\
&
\left\langle
\exp{\left( i \pi \sum_{l = i}^{k-1} S_{l,2}^z \right)}
\right\rangle_2 \times
\left\langle
\exp{\left( i \pi \sum_{l = i}^{k-1} S_{l,3}^z \right)}
\right\rangle_3
+ \ldots,
\end{align}
where $\langle \rangle_a$ is the expectation value with
respect to the groundstate of chain $a$.
While there are $3 \times 3 =9$ terms, each one of them
contains at least one factor of
\begin{align}
\left\langle
\exp{\left( i \pi \sum_{l = i}^{k-1} S_{l,a}^z \right)}
\right\rangle_a.
\label{eq:disorder-chain}
\end{align}
This is nothing but the disorder operator for the single chain,
introduced in Ref.~\onlinecite{RiseFall} and discussed
in Sec.~\ref{subsec: Stringorder}.
It vanishes because the ground state of each chain
in the Haldane phase.
As a consequence, the string order parameter~\eqref{eq:string-order1}
for the tube also vanishes, apparently implying that
the hidden \ZZ\ symmetry is unbroken.

On the other hand, the disorder operator~\eqref{eq:disorder} also
vanishes because it is simply a product of the disorder operators
for the three chains. 
This rather suggests that the hidden \ZZ\ symmetry is spontaneously
broken, in agreement with the edge state analysis.

The resolution of this apparent contradiction is as follows.
The hidden \ZZ\ symmetry is indeed broken spontaneously
even at the decoupling point $J_{\perp}=0$.
However, the string order
parameters~\eqref{eq:string-order1} and \eqref{eq:string-order2},
which are transformed to the ferromagnetic order in
$J^{z,x}$ by the non-local unitary transformation~\eqref{eq:transform},
are not ``good'' order parameters to detect the symmetry
breaking near the decoupling point.

In order to detect the hidden \ZZ\ symmery breaking
around the decoupling point, we introduce the following
variation of the string order parameter:
\begin{equation}
\langle \mathcal{O}^{zzz} \rangle = \lim_{|k-i| \rightarrow \infty}
\left \langle S^z_{i,1} S^z_{i,2} S^z_{i,3}
e^{ i \pi \sum_{l = i}^{k-1} J_l^z}
S^z_{k,1} S^z_{k,2} S^z_{k,3}
\right \rangle  .
\label{eq:product-string}
\end{equation}
The nonlocal transformation~\eqref{eq:transform} maps this
order parameter to a local order parameter
\begin{equation}
\langle V \mathcal{O}^{zzz} V^{-1} \rangle
= \lim_{|k-i| \rightarrow \infty}
\left \langle S^z_{i,1} S^z_{i,2} S^z_{i,3}
S^z_{k,1} S^z_{k,2} S^z_{k,3}
\right \rangle  .
\label{eq:VOzzzV}
\end{equation}
This does measure spontaneous breaking of the \ZZ\ symmetry
because $S^z_{i,1} S^z_{i,2} S^z_{i,3}$ is odd under
the global $\pi$-rotation about $x$ axis.

At the decoupled point $J_\perp =0$, Eq.~\eqref{eq:product-string}
reduces to the product of
the standard string order parameter~\eqref{eq:string_chain}
for the independent chains.
Since Eq.~\eqref{eq:string_chain} is non-vanishing in the
groundstate of each chain, the ``product'' string order
parameter~\eqref{eq:product-string} is also non-vanishing
in the tube at the decoupling point.
Therefore, the hidden \ZZ\ symmetry is indeed spontaneously
broken even at the decoupled point $J_\perp = 0$,
although the string order
parameters~\eqref{eq:string-order1} and \eqref{eq:string-order2}
cannot detect the symmetry breaking.
The edge spin $S_b=1/2$ in the weak coupling limit,
as well as in the strong coupling limit with anisotropy
$\alpha \lesssim 0.5$, can be understood as a consequence
of the hidden \ZZ\ symmetry breaking.
Thus no phase transition is expected between these two
regions, as both of them would belong to the
hidden \ZZ\ symmetry broken phase.

We note that the ``product'' string order
parameter~\eqref{eq:product-string}
is very similar
to the $\mathcal{O}_4$ defined in Eq.~(19) of Ref.~\onlinecite{Todo}.
However, there is a crucial difference between the
two-leg $S=1$ ladder case studied in Ref.~\onlinecite{Todo}
and the three-leg ladder/tube case discussed in this paper.
In the present case, the ``product'' string order parameter
detects spontaneous breaking of the hidden \ZZ\ symmetry,
thanks to the relation eq.~\eqref{eq:VOzzzV}.
However, in the two-leg ladder case, we find:
\begin{equation}
 V \mathcal{O}_4 V^{-1}  = \mathcal{O}_4 .
\end{equation}
Thus $\mathcal{O}_4$ introduced in Ref.~\onlinecite{Todo}
does {\em not} detect hidden \ZZ\ symmetry breaking.

In general, the product of string order parameters of each
chain is an order parameter for the hidden \ZZ\ symmetry
breaking in integer-spin ladder/tube with an {\em odd} number
of legs, but not with an {\em even} number of legs.
As a consequence, the hidden \ZZ\ symmetry (defined with
respect to Eq.~\eqref{eq:transform}) is broken
in the weak rung coupling limit of the odd-leg ladder/tube
with an odd integer spin, but remains unbroken in the same limit
if either the spin or the number of legs is even.

\subsection{Conjectured phase diagrams for $S = 1$ and $S = 2$ }

\begin{figure}[!h]
\begin{center}
\includegraphics*[width=7cm]{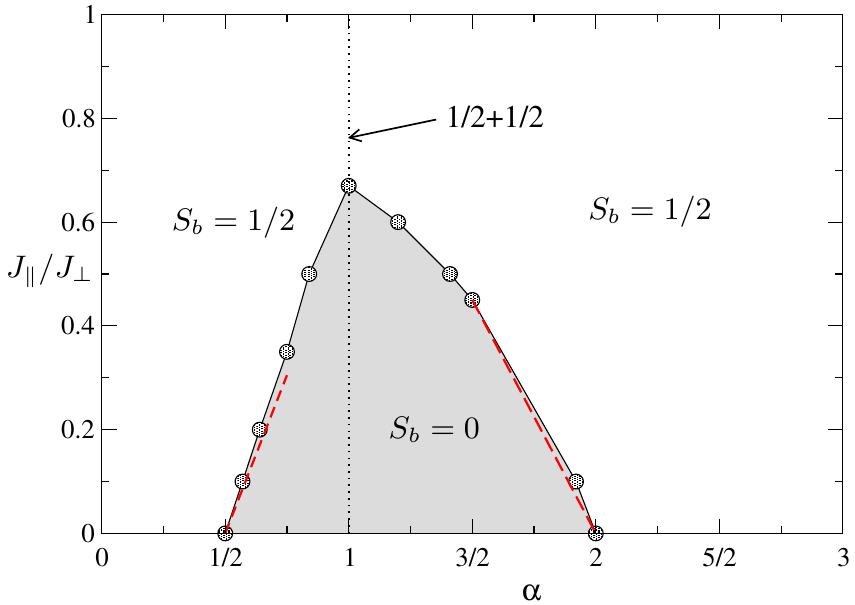}
\caption{(color online) Numerical phase diagram for the  spin $1$ tube obtained with DMRG (see part~\ref{subsec: DMRG}). Both phases
can be distinguished according to the string order parameters $\mathcal{O}^z$ (see Eq.(\ref{eq:string-order2}) or (\ref{eq:product-string})) and part \ref{subsubsec: stringorder}). 
The label $S_b$ for the different phases stands for the value of the spin of the boundary state for OBC. The transition between the $S_b=0$ and $S_b=1/2$ regime is analyzed in more detail in this paper and turns out to be first order. Phase boundaries obtained from an effective model (see part~\ref{subsec: Effectivemodel}) are also shown with dashed red lines  and agree quite well with numerical results.
}
\label{PhasediagS=1}
\end{center}
\end{figure}

Let us now discuss the phase diagram.
Here we propose the simplest phase diagram consistent
with our analyses in the previous subsections.
In Figure \ref{PhasediagS=1}, we show the conjectured/numerical 
phase diagram for a $S=1$ spin tube as a function
of $J_{\parallel} $ and $\alpha$.
In one of the two phases,
the hidden \ZZ\ symmetry is spontaneously broken and
the edge state with $S_b=1/2$ appears.
Although there are more edge state degeneracy on the
special lines $\alpha=1$ and $J_\perp=0$,
these lines are also a part of the broken hidden \ZZ\ symmetry
phase.

This phase diagram is also confirmed by numerical simulations;
the details will be given in Sec.~\ref{subsec: DMRG}.

\begin{figure}[!h]
\begin{center}
\includegraphics*[width=7cm]{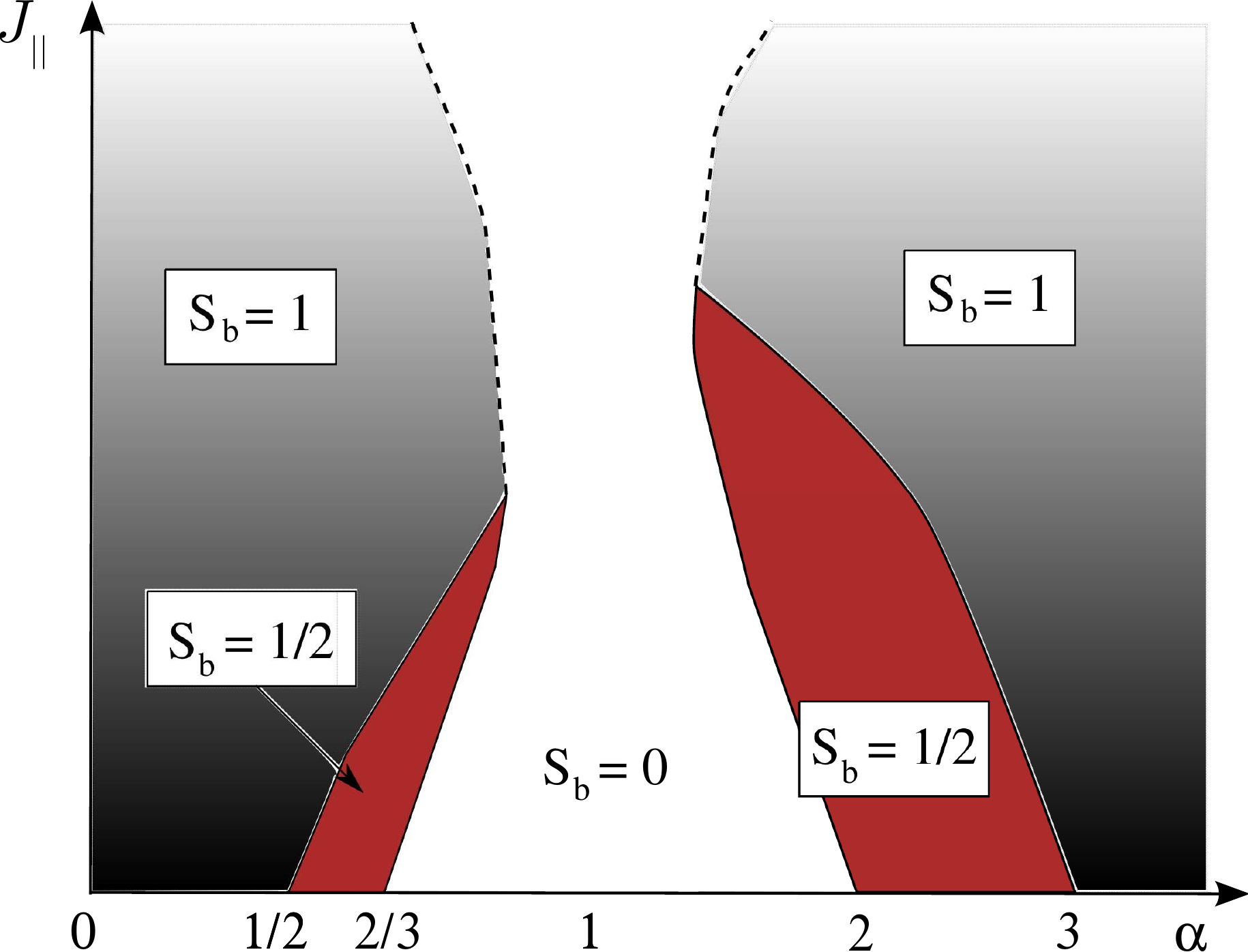}
\caption{(color online) Sketch of the phase diagram for the $S =2$ spin tube. Again, the label $S_b$ for the different phases stands for the value of the spin of the boundary state. The $S_b=1/2$ phases (light red) must be separated from the others with real phase transitions (full lines). On the other hand, there is no reason to expect a real phase transition between the regions $S_b=0$ and $S_b=2$ so we have plotted these boundaries with dash-dotted lines.}
\label{fig:PhasediagS=2}
\end{center}
\end{figure}

We can also conjecture the phase diagram for the $S=2$ spin tube,
as in figure~\ref{fig:PhasediagS=2}.
Based on the strong coupling
analysis, we in principle expect three different phases: a singlet phase
with no edge states, centered around $\alpha = 1$, one phase with
$1/2$ boundary states, and two phases with $S_b=1$ boundary spins.
We can also discuss the weak rung coupling limit
$J_\perp \rightarrow 0$ in terms of edge states, by
repeating the calculation of the
decoupled triangle of the preceding section but by this time reasoning
on the edge $S=1$ spins of the three $S=2$ chains.
We conclude, at the decoupling point $J_\perp =0$, that
there are no edge states for $0.5 < \alpha < 2$ and that there should
be a single spin $S_b=1$ at the boundaries for any other value of the
anisotropy parameter.

The edge state with $S_b=1/2$ would correspond to a spontaneous symmetry
breaking of the hidden \ZZ\ symmetry, which clearly
characterizes a distinct phase from those with $S_b=0,1$.
The phase with $S_b=1$ also appears to be different from
the $S_b=0$, concerning the edge state.
However, there is in principle no way to distinguish the phases
with $S_b=0$ and $S_b=1$.
This is suggested by the fact that hidden \ZZ\ symmetry is unbroken
in the $S=2$ Haldane ``phase''.\cite{Oshikawa}
In fact, it was pointed out recently in Ref.~\onlinecite{Pollmann09a}
that the $S=2$ Haldane ``phase'' with $S_b=1$ is adiabatically
connected to a trivial phase with $S_b=0$.

In our problem, the $S_b=0$ and $S_b=1$ ``phases'' are certainly
adiabatically connected at the decoupled point $J_\perp=0$,
where the system is just a collection of three $S=2$ chains
with the Haldane gap. With an infinitesimal coupling $J_\perp$,
the gap should not close for any value of $\alpha$.
Therefore we expect that $S_b=0$ and $S_b=1$ ``phases''
actually belong to a single phase in which the hidden \ZZ\ symmetry
is unbroken.

\section{\label{sec: NLSM}
The large $S$ limit: non Linear $\sigma$ model}
\subsection{Long wavelength description of the spin tube}
Having first examined the system from the strong coupling perspective,
we now shift to the examination of the large-$S$ approaches whose
greatest achievements culminate with the non linear sigma model. The
latter has shown to be particularly important in order to distinguish
the nature of the ground state, the low energy excitations and the
possible critical points of an
antiferromagnet~\cite{Manousakis,CHN,Johannesson}. It thus proves
valuable to conduct such a study here to complete our previous
analysis. The NL$\sigma$M can be derived from
the Heisenberg model when the spin $S$ is large. In
principle, it does not make any distinction between integer and
half-integer spins, as $S$ is just one between multiple parameters
allowed to flow continuously to their renormalized values at long
wavelength. However, the parity of the spin profoundly influences the
value of the Berry phase, a purely quantum quantity originating from
non zero overlaps between coherent states and entering the NL$\sigma$M
action. As shown by Haldane~\cite{Haldane1983}, the value of the Berry
phase eventually governs the properties of the system in the infrared
limit.

\begin{figure}
 \begin{center}
\includegraphics*[width=9cm]{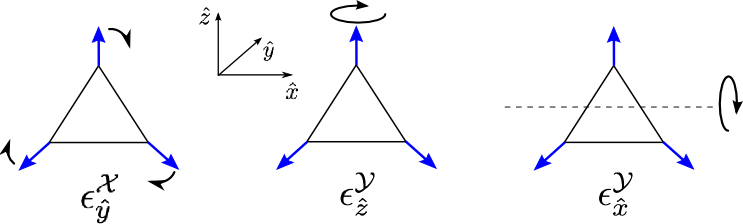}
\caption{(color online) The low energy modes deduced from the linear spin wave theory are connected to the three rotations of the initial triad around the axis $\hat{x}, \hat{y}, \hat{z}$. When propagating to the tube, they underpin slow twists of the original structure which asymptotically cost no energy at long wavelengths.}
\label{fig:modes}
\end{center}
\end{figure}

Although the spin-wave expansion cannot make good quantitative
predictions on a magnetic model in one dimension, it is useful to carry this
analysis in order to identify the low-energy, long-wavelength degrees
of freedom in the spin tube. These degrees of freedom will help later
on to construct a well-defined order parameter for the
NL$\sigma$M. A standard spin wave analysis shows that there
are three low energy modes depicted in Figure (\ref{fig:modes}) and
their canonical conjugate.
With the expression of the low-energy modes
$\eps^{\mathcal{ X}}_{\hat{y}}$,
$\eps^{\mathcal{ Y}}_{\hat{z}}$ and
$\eps^{ \mathcal{Y}}_{\hat{x}}$, and
their respective conjugate modes
$\eps^{ \mathcal{Y}}_{\hat{y}}$,
$\eps^{ \mathcal{X}}_{\hat{z}}$ and
$\eps^{ \mathcal{X}}_{\hat{x}}$,
we can reexpress the slowly varying spin degrees of freedom.
The spin operators can be rewritten in a compact form by introducing
an infinitesimal $SO(3)$ rotation matrix $\R_i = \exp \left (i
  \mathbf{m}_i \hat{\mathbf{J}} \right)$ with $\hat{\mathbf{J}}$
standing for the generator of $SO(3)$, and a vector $\mathbf{L}_i$ if
we identify:
\begin{equation*}
\mathbf{L} = \Vectthree{\frac{-2\alpha}{\sqrt{4\alpha^2+2}} \eps^\mathcal{X}_{\hat{x}}}{\frac{\eps^\mathcal{Y}_{\hat{y}}}{3}}{-\frac{\eps^\mathcal{X}_{\hat{z}}}{\sqrt{2-\frac{1}{2\alpha^2}}}} \; \; \;
\mathbf{m} = i \Vectthree{\frac{2\alpha}{\sqrt{4\alpha^2+2}} \eps^\mathcal{Y}_{\hat{x}}}{\frac{\eps^\mathcal{X}_{\hat{y}}}{3}}{\frac{\eps^\mathcal{Y}_{\hat{z}}}{\sqrt{2-\frac{1}{2\alpha^2}}}}.
\end{equation*}
The original spin operators read very simply:
\begin{equation}
\mathbf{S}_{i,a} =(-1)^i \R_i\n_a +(\mathbf{L}_i - (\mathbf{L}_i \cdot \n_a)\n_a).
\label{eq:ansatz}
\end{equation}
where the vectors $\n_a$ are given in (\ref{eq:coplanar}).
The number of degrees of freedom on the left and on the right hand side of this equation is the same. Thus, this operation can be regarded as a simple change of variables, but if one is interested in a long wavelength limit, the new variables $\R_i$ and $\mathbf{L}_i$ are the most adapted to describe the physics of the anisotropic spin tube.

 \subsection{Derivation of the NL$\sigma$M}

 After having derived the low energy modes of the theory, we can now
 focus on the construction of the NLsM.  The first point is to define
 a suitable local order parameter. In the case of a coplanar
 configuration, the role of the order parameter will be played by the
 $SO(3)$ rotation matrix. The underlying idea behind the construction
 of the NLsM is then to consider the system in its symmetry-broken
 phase and to take into account quantum fluctuations around the
 direction of the order parameter. It is actually not necessary that
 the system admits a broken-symmetry phase (it has none in $1d$), it
 just needs to be at least ordered locally, which is the case in the
 weak rung coupling limit $J_\parallel \gg J_\perp$.
 The continuum limit
 is then reached within a Hamiltonian formalism or a path integral
 (lagrangian) approach~\cite{Fradkin}. One of the greatest
 possibilities given by the NLsM is that it can be investigated with
 renormalization group techniques for which one can determine the
 nature of the possible fixed points governing the physics at long
 wavelengths. The quantum NLsM for triangular geometries and its RG
 analysis have already been extensively studied by Azaria \textit{et
   al.}~\cite{Azaria,Mouhanna}. We shall refer later to their work for
 the characterization of the spectrum. For now, our starting point
 will be different. We will build the NLsM in the lagrangian
 formalism, following the construction of Dombre and
 Read~\cite{Dombre}. This path integral approach is particularly
 illustrative regarding the construction of the Berry phases. These
 complex terms will play a major role in our analysis of the quantum
 spin tube.

 In the long wavelength limit, the fine scale of the lattice becomes
 irrelevant and the lattice spacing $\lambda$ can be taken to
 zero. 
To  obtain a well-defined continuum limit, we need to choose a local
order-parameter field which has a smooth spatial variation on the
scale of $\lambda$. To do so, we make use of the following
ansatz~\cite{Dombre}:
\begin{equation}
\mathbf{S}_{i,a}(t) =
 S\frac{ \R_i(t) \left( (-1)^i \mathbf{n}_a+ \lambda \mathbf{L}_i(t) \right)}{
\vert \R_i(t) \left( (-1)^i \mathbf{n}_a+  \lambda \mathbf{L}_i(t) \right) \vert },
\end{equation}
where the fields $\R$ and $\mathbf{L}$ depend
on the time $t$ and the lattice coordinate $i$.
This is identical to (\ref{eq:ansatz}) at first order in
$\lambda$. Here however, to make a coherent calculation, we will need to
keep the development up to \textit{second order} in $\lambda$:
\begin{eqnarray}
\mathbf{S}_{i,a} &\approx&  S \R_i\left( (-1)^i \n_a +
\lambda ( \mathbf{L}_i - (\n_a . \mathbf{L}_i)\n_a) \right.
\label{eq:ansatz2}
 \\
&+& \left. (-1)^i \lambda^2
 \left(\left( -\frac{\mathbf{L}_i^2}{2} +
	\frac{3}{2} (\n_a . \mathbf{L}_i)^2
 \right) \n_a - (\n_a . \mathbf{L}_i)
\mathbf{L}_i \right) \right). \nonumber
\end{eqnarray}
In particular, the square of the magnetization of the whole triad is given by:
\begin{align*}
& \left( \sum_a \S_{i,a} \right)^2 \approx
S^2 \left( \frac{\alpha-1}{\alpha} \right)^2 +
9\lambda^2 S^2 (T \mathbf{L}_i)^2 +\\ 
& \lambda^2 S^2 \left( \frac{\alpha-1}{\alpha} \right)^2
\left( -\mathbf{L}_i^2 - 2{L_i^z}^2 + 3\frac{\alpha}{1-\alpha}
\sum_a (\n_a . \mathbf{L}_i)^2( \n_a . \hat{z}) \right) ,
\end{align*}
where $T_{\alpha \beta} = \delta_{\alpha \beta} - \frac{1}{3}\sum_a
n_{a\alpha} n_{a\beta}$,
and $(T \mathbf{L})_{\alpha} = \sum_\beta T_{\alpha \beta} L_\beta$.
We set again $J_\parallel = J_\perp = 1$
for simplicity and concentrate on the effects of the anisotropy
parameter $\alpha$. The action we wish to estimate is~\cite{Fradkin} :
\begin{align}
S_{{\rm NL}\sigma{\rm M}} &= S_{BP} + S_H \nonumber \\
&= i\sum_{i,a} \omega[\S_{i,a}]  - \int \ud \tau H[\S_{i,a}],
\label{eq:Action}
\end{align}
$\tau$ denoting imaginary time. The first part of the action is the
Berry phase term. It measures the total area covered by each of the
spins $\S_{i,a}$ on the sphere of radius $S$.
Up to second order, the Berry phase term reads:
\begin{equation}
S_{BP} = iS \sum_{i,a}(-1)^i \omega[\R_i \n_a]  +
i 3  S\int \ud \tau \ud x \;
\left( T \mathbf{L}_i \right) \cdot \mathbf{V}, 
\label{eq:Berrytot}
\end{equation}
where $V_{\alpha} = -\frac{1}{2} \eps^{\alpha \beta \gamma} (\R^{-1}\partial_\tau \R)_{\beta \gamma}$ and $\eps^{\alpha \beta \gamma}$ is the totally antisymmetric tensor. The first member of the right-hand side,
\begin{equation}
S^\prime_{BP} =   iS \sum_{i,a} (-1)^i\omega[\R_i \n_a] 
\end{equation}
takes a particular significance when one allows for the possibility of
singularities in the action. These singularities are naturally present
in the system because we start from a lattice description with
discrete variables and not fields. However, we wish not to take them
into account now and we will let apart this term for the moment. We
will reconsider it when evaluating the role of the topological defects
in the theory.

The second member of Eq.~(\eqref{eq:Action}) is nothing but the
Hamiltonian~(\ref{eq:Hamiltonian}) where the quantum spin operators
have been replaced by the ansatz (\ref{eq:ansatz2}).
The Hamiltonian part, at second order in $\lambda$, reads:
\begin{align*}
S_H &= \int \ud x \ud \tau \left( - \frac{1}{\lambda} S_{\rm triad} - 6 \lambda S^2T \mathbf{L}(x,\tau) \cdot \mathbf{L}(x,\tau) \right.\\
&+ S \left. \Tr[P(\R(x,\tau)^{-1} \partial_x \R(x,\tau))^2] \right).
\end{align*}
with:
\begin{equation*}
S_{\rm triad} = \frac{1}{2} \left(\sum_a \S_{i,a} \right)^2 - \frac{1-\alpha}{2} \left(\S_{i,1} +\S_{i,2}\right)^2. \\
\end{equation*}

Since the action is quadratic in the field $\mathbf{L}(x,\tau)$, we can integrate the field out and finally express the action solely in terms of the $SO(3)$ matrix field $\R(x,t)$: 
\begin{eqnarray}
S_{\rm NL\sigma M} &=&  S \int \ud x \ud \tau \,  \left( \Tr [ P (\R(x,\tau)^{-1} \partial_x \R(x,\tau))^2 ] \right. \nonumber \\
&+& \left. \Tr [ Q (\R(x,\tau)^{-1} \partial_\tau \R(x,\tau))^2 ] \right) + iS^\prime_{BP}. \label{eq:NLSM2} 
\end{eqnarray}
Here $P$ and $Q$ are diagonal matrices whose expression is better
given by the spin-stiffness and susceptibility tensors~\cite{Azaria}:
\begin{equation}
\chi_{\alpha \beta} = - \Tr(Q t_\alpha t_\beta),
\hspace{5mm} \rho_{\alpha \beta} = -\Tr(P t_\alpha t_\beta),
\end{equation}
with $\chi_{\alpha \beta} = \chi_\alpha \delta_{\alpha \beta}$, $\rho_{\alpha \beta} = \rho_\alpha \delta_{\alpha \beta}$ and:
\begin{eqnarray}
\chi_1 &=& \frac{S(1+2\alpha^2)^2}{\lambda \alpha(1+4\alpha(3+\alpha+4\alpha^2))}, \;
\chi_2 = \frac{9S\alpha}{\lambda( 2+8\alpha(4+\alpha))}, \nonumber \\
\chi_3 &=& \frac{S}{\lambda} \left( \frac{1}{\alpha} - \frac{8}{-1+4\alpha(2+\alpha)} \right), \nonumber \\
\rho_1 &=& \lambda S\left(1+\frac{1}{2\alpha^2} \right), \; \rho_2 = 3\lambda S, \; \rho_3 = \lambda S \left( 2-\frac{1}{2\alpha^2} \right). 
\label{eq:couplings}
\end{eqnarray}
Here $t_\alpha$ are the generators of the $SO(3)$ group.
Finally, we introduce the fields
$\R(x,\tau)^{-1} \partial_\mu \R(x,\tau) = \omega^\alpha_\mu t_\alpha$.
The action reads~\cite{note-velocity}: 
\begin{equation}
S_{\rm NL\sigma M} =  \frac{S}{2} \int_0^\infty \ud \tau \ud x \left( \chi_{\alpha \beta} \omega_0^\alpha \omega_0^\beta + \rho_{\alpha \beta} \omega_x^\alpha \omega_x^\beta \right) + iS^\prime_{BP}, 
\label{eq:NLSM3} 
\end{equation}

\subsection{\label{subsec: Bare}
Bare analysis of the NL$\sigma$M.}

The two formulations of the action, eqs. (\ref{eq:NLSM2}) and
(\ref{eq:NLSM3}), are valid for all $\alpha \geq 0.5$. In particular,
we should be able to recover the isotropic limit $\alpha = 1$, and the
unfrustrated cases $\alpha = 0.5$ and $\alpha \rightarrow \infty$. For
$\alpha =1$, one notes that $\rho_1 = \rho_3 \neq \rho_2$ and $\chi_1
= \chi_3 \neq \chi_2$. In the language of the $SO(3)$ matrices, this
translates into an additional $SO(2)$ global right symmetry of the
action: $\R \rightarrow \R U_R$. This symmetry is reminiscent of the
discrete $C_{3v}$ symmetry of the triangle for $\alpha = 1$. Since the
configurations of fields classically minimizing the action also
possesses a $SO(2)$ symmetry, this model is referred to as the
$SO(3)\times SO(2)/SO(2)$ NL$\sigma$M~\cite{Delduc,Azaria}. When
$\alpha = 0.5$, one finds $\rho_1 = \rho_2$, $\chi_1 = \chi_2$ and
$\chi_3 = \rho_3 = 0$ and recovers the description of the collinear
antiferromagnet in terms of a $O(3)/O(2)$ NL$\sigma$M.


There is another, third representation of the action
(\ref{eq:NLSM2})-(\ref{eq:NLSM3}) that nicely illustrates the effect
of the anisotropy parameter $\alpha$. Remembering that a $SO(3)$
matrix is nothing more but a set of three orthonormal vectors:
$(\mathbf{e}_b)_a = \R_{ab}$, we can use the fact that $\mathbf{e}_2 =
\mathbf{e}_3 \times \mathbf{e}_1$ to rewrite the {\it bare} action in terms of
two orthonormal unit vectors:
\begin{eqnarray}
S_{\rm NL\sigma M} &=& S_1+S_3+ S_{\textrm{coupling}} +S^\prime_{BP} , \\
S_a &=& \frac{1}{2\tilde{g}_a} \int \ud x \ud \tau
 \left(\tilde{c}_a(\partial_\tau \mathbf{e}_a)^2+
  \frac{1}{\tilde{c}_a}(\partial_x \mathbf{e}_a)^2  \right), \nonumber \\
S_{\textrm{coupling}} &=& - \frac{\kappa}{2}  \int \ud x \ud \tau
 (\mathbf{e}_3 \cdot \partial_\tau \mathbf{e}_1)^2 ,
\nonumber
\label{eq:NLSM4}
\end{eqnarray}
where the new constants can be easily expressed as a function of the
spin-stiffness and susceptibility tensors.
The behavior of the different couplings
as a function of $\alpha$ are easily obtained from the expression of the
spin-stiffness and susceptibility tensors. An important point is
that for $\alpha \to 0.5$, $g_1 \to \infty$ and $\kappa \to 0$ {\it
i.e.} when $\alpha = 0.5$ the $\mathbf{e}_1$ field becomes a spurious
degree of freedom with null stiffness! This is consistent with the
collinear picture in the range $0 \leq \alpha \leq 0.5$. Such a model is
in fact well described by the fluctuations of a single unit vector
$\mathbf{e}_3$.

   
\subsection{\label{subsec: RG}
Renormalization group analysis of the NL$\sigma$M.}

The above bare analysis of the NL$\sigma$M is insufficient to describe
properly the behavior of the spin tube in the infrared limit. As it is
well known from the study of the quantum spin chain, quantum
fluctuations always renormalize the parameters entering a NL$\sigma$M
action and eventually drive the system into a quantum disordered state
in $1d$~\cite{Polyakov}. Thus, in order to understand the properties
of the system at long wavelength, one must perform a renormalization
group analysis to determine how do the different coupling constants
(\ref{eq:couplings}) renormalize. We are going to use the results for
the one-loop RG equations\cite{Azaria,Mouhanna}, starting with the set of bare
couplings (\ref{eq:couplings}). To make the distinction between
isotropic and anisotropic cases, we introduce the two anisotropy
parameters $\alpha_2 = 1 - \rho_2/\rho_1$ and $\alpha_3 =
1-\rho_3/\rho_1$ and the coupling $g = 2/\rho_1$, the latter playing
the role of an effective coupling constant. The set of couplings
$\gamma = \{c_1,c_2,c_3,\alpha_2,\alpha_3,g \}$ obeys the general RG
equations:

\begin{equation}
\frac{\partial \gamma}{\partial l}= - \beta(\{ \gamma \})
\end{equation}

We have integrated numerically these equations for different values of $\alpha$ ranging from $\alpha = 0.55$ to $\alpha = 0.95$ (similar behaviors were also observed for values above $\alpha = 1$). 
The numerical integration of the RG equations yields the unambiguous
result that the symmetry is dynamically \textit{enlarged} in the
infrared limit, similarly to the higher dimensional cases. For any
value of $\alpha > 0.5$, the spin wave velocities renormalize to the
same value $c_1^\star = c_2^\star = c_3^\star$ while the two
anisotropy parameters $\alpha_2$ and $\alpha_3$ fall to zero. The
symmetry of the model in the long-wavelength limit is therefore $SO(3)
\times SO(3) / SO(3) \approx SO(4)/SO(3)$. The coupling $g$ diverges,
as one could have expected in one dimension. Hence, there seems to be
no qualitative differences between the cases $\alpha = 1$ and $\alpha
> 0.5, \alpha \neq 1$ at the one loop level, suggesting that a
deviation from the point $\alpha =1$ is an irrelevant perturbation.
However, we have not taken into account so far the role of the Berry
phase term, which as we are going to see plays an important role when
$\alpha \neq 1$.

\section{\label{sec: Berry} Berry phases and Instantons}
\subsection{Instantons in SO(3) NL$\sigma$M}
The continuous part of the sigma model does not make any distinction
between the integer and the half-integer quantum spin tube. In fact,
the preceding RG equations suggest that the model is gapped in both
cases, and admits a unique ground state for any $\alpha$. Nonetheless,
the DMRG data show unambiguously a dimerization of the ground state of
the spin tube for $S = \frac{1}{2}$ at $\alpha =
1$~\cite{Sakai,Nishimoto}. Analogously, the Majumdar-Gosh model, whose
NL$\sigma$M also has the $SO(3)\times SO(2)/SO(2)$ symmetry, is
dimerized~\cite{Majumdar,Rao}. In a single spin chain, the difference
between integer and half-integer spins can be explained in the
NL$\sigma$M by the presence of a topological term in the
action~\cite{Haldane1983}. Here, such a term is absent because of the
triviality of the second homotopy group of the $SO(3)$
manifold~\cite{Dombre,Mermin}:
\begin{equation*}
\pi_2(SO(3)) = 0
\end{equation*}
However, only continuous space-time configurations of the field $\R(x,\tau)$ have been considered up to now. In fact, there also exist configurations containing vortices with singular cores. These defects originate from the non trivial first homotopy group of $SO(3)$:
 \begin{equation}
\pi_1(SO(3)) = \mathbf{Z}_2.
\label{eq:Z2}
\end{equation}
For classical antiferromagnets on the triangular lattice, these
vortices are argued to be the driving force of a phase
transition~\cite{Miyashita}.

In quantum systems, topological defects
radically affect the behavior of the disordered phases of the
$O(3)/O(2)$ NL$\sigma$M in $2d$~\cite{Haldane1988}, leading the system
to dimerization, and of the $SO(3)\times SO(2)/SO(2)$ NL$\sigma$M in
$1d$~\cite{Rao}. The specificity of our system is that the $SO(2)$
symmetry is, at least at the bare level, no longer present when
$\alpha \neq 1$. Thus, we would like to investigate the conjugate
action of the topological defects, also known as
instantons~\cite{Polyakov}, and of the anisotropy in the spin
tube. For integer $S$, we will see that the presence of the
topological defects gives rise to the emergence of $2S$ peculiar values
of $\alpha$, that we could associate with the critical points
determined from the strong coupling approach.

We would like to review first the nature of the instantons in our
system. Instantons are topological defects associated to the symmetry
group of the order parameter. The $SO(3)$ group manifold is isomorphic
to a ball of radius $\pi$ in three dimensions whose diametrically
opposite points on the surface are identified. One can associate to a
rotation around an axis $\n$ by an angle $\theta$, the vector $\theta
\n$ with $\theta \in [-\pi,\pi]$. The redundancy between two opposite
points on the surface of the sphere stems from the identification
between a rotation about an axis $\mathbf{n}$ of angle $\pi$ and the
rotation about the same axis of angle $-\pi$. It is then clear that
the $SO(3)$ manifold is non simply connected, with the ensemble of
closed path in $SO(3)$ dividing into two classes: one containing the
loops shrinkable to a point, the other ones containing strings joining
two opposite points of the ball. This is equivalent to say that 
the first homotopy group of the
$SO(3)$ manifold is given by
(\ref{eq:Z2}). Considering the evolution of a matrix $\R(x)$ through
space, an element of the non trivial class is:
\begin{equation}
\R(t) = \Matthree{\cos \theta(x)}{\sin \theta(x)} {0}{-\sin \theta(x)}{\cos \theta(x)} {0}{0}{0}{1},
\end{equation}
with $\theta(x = 0) = 0$ and $\theta(x = L) = 2\pi$, where $L$ is the size of the system. Conversely, the trivial class will consist of matrices which stay close to the identity matrix at all positions. 

Turning back to our $1+1$ dimensional problem, suppose now we start
from a configuration in the trivial sector with all $\mathbf{e}_1$
vectors pointing up and all $\mathbf{e}_2$ vectors pointing right,
where again $\R_{\alpha\beta} = (\mathbf{e}_\alpha)_\beta$ (see Fig
\ref{fig:instantons}). If nothing "sudden'' happens, \textit{i.e.} if
the time evolution process is sufficiently smooth, the chain should
visit other configurations but stay in the trivial topological
class. However, it is also possible that some non trivial
configurations arise during time evolution that will connect the two
classes of path. These are the \textit{instantons}. A pair of
instanton (going from the trivial to the non-trivial class) and
anti-instanton (i.e the opposite) is represented on Fig.~\ref{fig:instantons}.  In the continuum, an instanton will appear as
a singularity. It is clear that such an event is unlikely to happen if
the tube is ordered. However, since the model is disordered at long
wavelengths, these events will eventually proliferate. Now, it may be
that the proliferation of instantons is constrained by the Berry phase
term. Here, we would like to calculate:

\begin{figure}
\begin{center}
\includegraphics*[width=5cm]{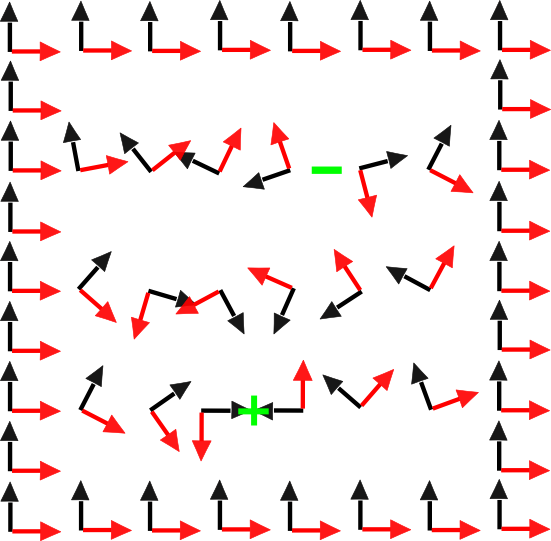}
\caption{(color online) Evolution of the spin tube from the two different topological
  sectors of $SO(3)$. Each triad is represented by two orthonormal
  vectors $\mathbf{e}_1(x,t)$ and $\mathbf{e}_2(x,t)$ that one can
  connect with the classical configuration of spins on each triangle
  $(\S_{i,1},\S_{i,2},\S_{i,3})$~\cite{Delduc}. In the continuum,
  $\mathbf{e}_1$ and $\mathbf{e}_2$ also stand for the first two
  vector columns of the rotation matrix $\R(x,t)$. Starting from a
  N\'eel configuration, the system tunnels to a non trivial
  configuration via an instanton event $(+)$. The system returns to
  the trivial configuration via an anti-instanton $(-)$.}
\label{fig:instantons}
\end{center}
\end{figure}

\begin{equation}
S^\prime_{BP} = i S\sum_{i,a} \omega[\R_i \n_a] (-1)^i.
\label{eq:Berry}
\end{equation}
which is the discrete part of the total Berry phase
(\ref{eq:Berrytot}) that we let apart.  For this purpose, we follow
Dombre and Read again and consider a first time path $\R(\tau)$
satisfying the closed boundary conditions and a second one $\R'(\tau)
= \R(\tau) + \delta \R(\tau)$ infinitesimally close to $\R(\tau)$. The
difference of Berry phases between the two paths can be easily
evaluated to be:
\begin{eqnarray*}
\delta S^\prime_{BP} &=& iS\int \ud \tau \sum_a (\delta \R \n_a) \cdot (\partial_\tau \R \n_a \times \R \n_a) \\
&=& -iS \int \ud \tau V_{\beta}  (\R^{-1} \delta \R)_{\beta \beta'} \left( \sum_a \n_{a\beta'} \right).
\end{eqnarray*}

\subsection{Isotropic case, $\alpha = 1$}
\begin{figure}
\begin{center}
\includegraphics*[width=5cm]{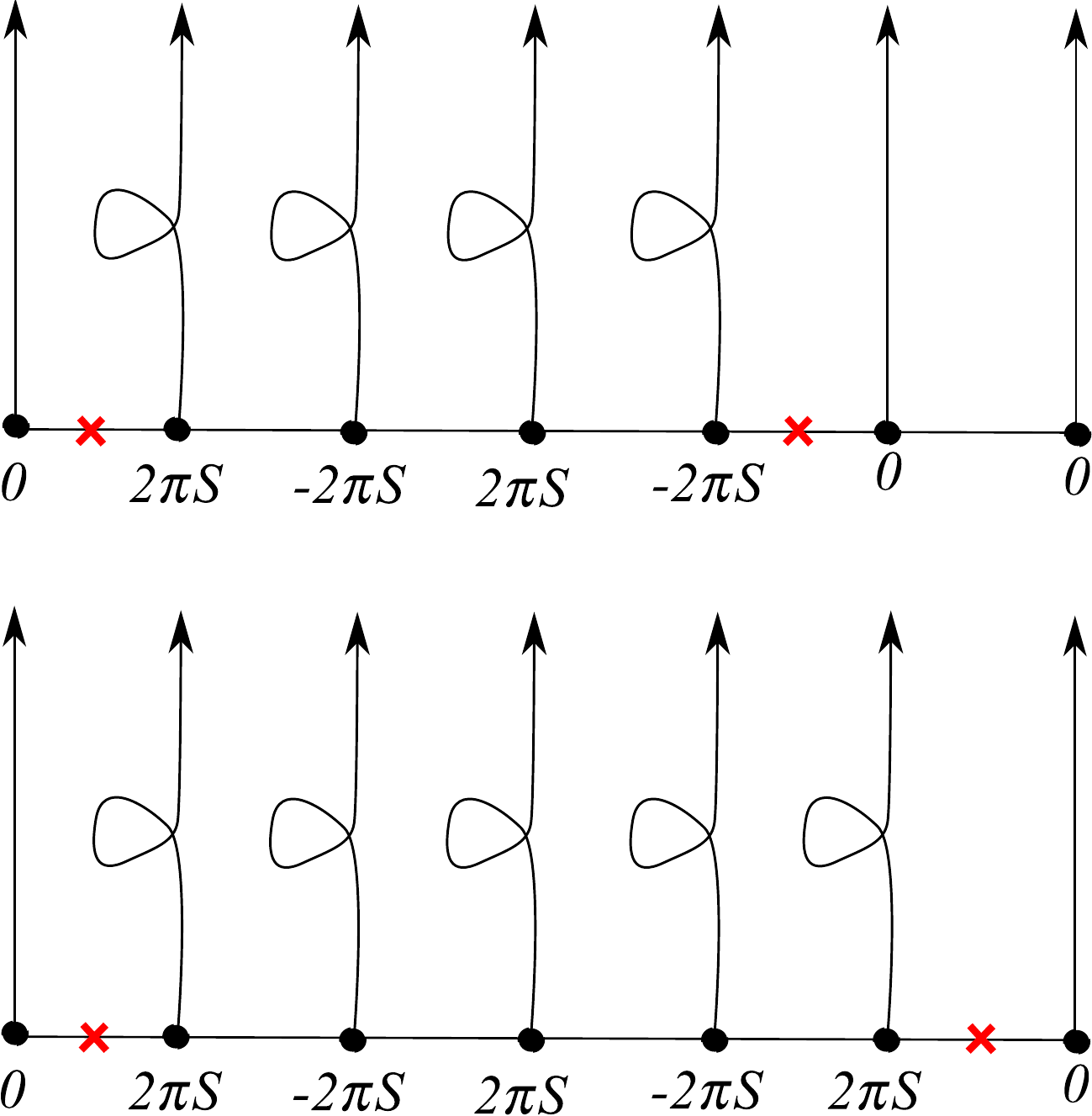}
\caption{(color online) Two space time configurations differing by the shift of one instanton. The two crosses represent the positions of the singularities. A straight line represents a trivial path in $SO(3)$ while a loop is a non-trivial path. The Berry phase associated with each loop is $\pm 2\pi S$ depending on the sublattice.}
\label{fig:instantons2}
\end{center}
\end{figure}

In the isotropic case, we have the important result that:
\begin{equation}
\delta S'_{BP} = 0
\end{equation}
and any smooth change in the history of $\R(\tau)$ will not change the
value of the Berry phase of the triad. Thus, this quantity can be used
to index the two classes of $\pi_1(SO(3))$, exactly like the hedgehog
number classifies the configurations of the spins in two
dimensions~\cite{Haldane1983,Fradkin}. Because the quantity is a
topological invariant, we just need to calculate it for one path
representing each class. For the trivial class, we can take the
identity matrix so that $\sum_a \omega[ \R(t) \n_a] = 0 \, [4\pi
S]$. For the non-trivial class, we can consider the rotation of the
triad around an arbitrary axis. In this case, the Berry phase of the
triad will be $2 \pi S \, [4 \pi S]$. So, the alternating sum
(\ref{eq:Berry}) reads:
\begin{equation}
S^\prime_{BP} = iS\sum_{i,a} 2 \pi  q_i (-1)^i,
\label{eq:Berry2}
\end{equation}
where $q_i = 0,1$ depending on which class the matrix $\R_i$ belongs to. Consequently, the total Berry phase will be $0$ or $2\pi S$ depending on the number of non-trivial paths. If $S$ is an integer, the Berry phase has no effect. But if $S$ is a half-integer, we see that there are two different values for $S_{BP}$, defining two different vacua. \\
To see the influence of the instantons on the system, we remember the
arguments of Rao and Sen~\cite{Rao}. An instanton is a discontinuity
in the Berry phase of two neighboring triads.  Because of closed
boundary conditions in the partition function, an instanton
necessarily comes with an anti-instanton. As we saw, the creation of
such a pair links the two vacua labelled by $S_{BP} = 0$ and $S_{BP} =
2\pi S$. The instantons are situated on the links of the lattice (as
they live on the plaquettes of the lattice in the (2+1)d case). A pair of
instanton, anti-instanton defines a string of a given size. If this
size is even, the Berry phase of the string is $0$; if it is odd, it
is $2\pi S$ (Fig \ref{fig:instantons2}). It is then easy to see that
if $S$ is half-integer, there will be destructive interferences
between paths with strings of different sizes. In particular, if we
shift an instanton by one lattice site, we expect the dynamical
contribution from the Hamiltonian to not change, but the Berry phase
to change by $2 \pi S$. For instance, the two paths of Figure
\ref{fig:instantons2} will contribute in the partition function:
\begin{equation}
Z = \ldots + (1+\textrm{e}^{2i\pi S})e^{S_{\mathrm{NL\sigma M}}} + \ldots
\end{equation} 

For half-integer spins, different instantons-anti-instantons
contributions are compensated by destructive interferences. Thus, the
two topological sectors $q = 0,1$ are non connected and we are left
with two degenerate ground states labeled by the two elements of
$\pi_1(SO(3))$. As shown by Read and Sachdev in a large $N$ analysis
of the (2+1)d Heisenberg model~\cite{SachdevRead1}, this kind of
destructive interferences between instantons leads to dimerization in
disordered phases. This seems to be the case here: the spin $1/2$
isotropic model is known to be dimerized by DMRG.

On the other hand, integer spins allow instanton events to proliferate
as all events come with the same phase. This makes that the two vacua
are well connected. This ``tunneling'' between vacua lifts the
degeneracy and the ground state is therefore unique.

\subsection{Anisotropic case, $\alpha \neq 1$}

For $\alpha \neq 1$, the difference in Berry phase between two matrices $\R$ and $\R + \delta \R$ belonging to the same topological class is:
\begin{eqnarray}
\delta S_{BP}^\prime &=& -iS \int \ud \tau \; V_{\beta}  (\R^{-1} \delta \R)_{\beta \beta'} \sum_a \n_{a\beta'}  \nonumber \\
&=& i\frac{S}{2} \frac{1-\alpha}{\alpha} \int \ud \tau \; \delta \mathbf{e}_3 \cdot ( \partial_\tau \mathbf{e}_3 \times \mathbf{e}_3 ).
\end{eqnarray}
In the anisotropic case, the Berry phase can no longer be used to classify topological classes. For example, the path contribution to the partition function from the two configurations drawn on figure \ref{fig:instantons2} is now:

\begin{eqnarray}
Z &=& ... + \textrm{e}^{2i\pi S L}\left( \textrm{e}^{ i\frac{S}{2} \frac{1-\alpha}{\alpha} \int_0^L \ud x \int \ud \tau \; \partial_x \mathbf{e}_3 \cdot ( \partial_\tau \mathbf{e}_3 \times \mathbf{e}_3 )} \right. \label{eq:paths} \\
&+& \left. \textrm{e}^{i2\pi S} \textrm{e}^{i\frac{S}{2} \frac{1-\alpha}{\alpha} \int_0^{L+1} \ud x \int \ud \tau \; \partial_x \mathbf{e}_3 \cdot ( \partial_\tau \mathbf{e}_3 \times \mathbf{e}_3)}\right)e^{S_{\mathrm{NL\sigma M}}} + ... \nonumber
\end{eqnarray}

Here, we have been careful to write the total Berry phase as a
continuous integral over the string. By doing so, we made the
approximation that the order parameter $\R(x,t)$ is sufficiently
smooth so that the derivatives $\partial_x \mathbf{e}_i$ are
well-defined. This development is valid if we stay in a given
topological sector of $SO(3)$. For a configuration with many
instantons, we should separate the contributions from the different
topological sectors and write it as a sum of integrals:

\begin{eqnarray}
  S_{BP}^\prime &=&  i\sum_{i,a} 2 \pi S q_i (-1)^i \label{eq:Berry3} \\
  &+& i\frac{S}{2} \frac{1-\alpha}{\alpha} \sum_{i = 0}^{P-1} \int_{x_i}^{x_{i+1}} \ud x \int \ud \tau \; \partial_x \mathbf{e}_3 \cdot ( \partial_\tau \mathbf{e}_3 \times \mathbf{e}_3 ),
  \nonumber 
\end{eqnarray}
where $\{x_1,\ldots ,x_i\}$ denotes the position of the $P$
instantons. Note that without instantons (i.e if $\R(x,t)$ is a smooth
field everywhere), we can regroup all the integrals into a single one
and this term identically vanishes:
\begin{eqnarray}
 & & \int_{- \infty}^{+\infty} \ud x \int \ud \tau \; \partial_x \mathbf{e}_3 \cdot ( \partial_\tau \mathbf{e}_3 \times \mathbf{e}_3 )  \label{eq:supequal}\\
 &=& \int \int \ud x  \ud \tau \;  \left[ \partial_\tau (\mathbf{e}_1 \cdot \partial_x \mathbf{e}_2) - \partial_x (\mathbf{e}_1 \cdot \partial_\tau \mathbf{e}_2) \right] = 0, \nonumber
\end{eqnarray}
given the periodic boundary conditions we imposed.

Let us finally recover some well-known result in the extreme limits $\alpha = 0.5$ and $\alpha \rightarrow \infty$. In this case, the symmetry of the order parameter reduces to $O(3)/O(2) \cong S^2$. There are no instantons in this case since $\pi_1(S^2) = 1$. It is then straightforward to show that (\ref{eq:Berry3}) reduces to:
\begin{equation*}
S_{BP}^{\rm tot \prime} = i\frac{S}{2} \int_{-\infty}^{+\infty} \ud x \int \ud \tau \; \partial_x \mathbf{e}_3 \cdot ( \partial_\tau \mathbf{e}_3 \times \mathbf{e}_3 ) = 2i \pi S n \; n \in \mathbb{Z}
\end{equation*}
We recall that having a non-trivial skyrmion number for a smooth space-time configuration of the vector field $\mathbf{e}_3$ requires discontinuities on
the field $\mathbf{e}_1$. However, as this last field gets zero stiffness for $\alpha = 0.5$ and decouples
we recover for the field $\mathbf{e}_3$ the form of the NL$\sigma$M with the correct topological term of a single chain of spins $S$ as we should. 

\subsubsection{Half-integer spins}
\label{sec:Berry}
Reiterating the argument that led us to the twofold degeneracy of the
ground state for $\alpha =1$ and half-integer spins, we find with
(\ref{eq:paths}) that the different instantons-anti-instantons
contributions do \textit{not} cancel out anymore. The tunneling process between 
the two topological
sectors is present and the topological degeneracy is
lifted. Consequently, we can make use of an important result for spin
chains, the Lieb-Schultz-Mattis theorem~\cite{LSM}, suggesting that
the system is in a \textit{gapless} phase. This theorem states that
spin Hamiltonians with local interactions and an half-integer spin per
unit cell like (\ref{eq:Hamiltonian}) either support gapless
excitations or have a ground-state degeneracy. Ruling out the
possibility of a degeneracy here tends to the scenario of a critical
behavior. This is indeed what appears in the the study of T. Sakai
\textit{et al} where, for $S=1/2$, the DMRG
data point at a preservation of the spin gap
only in a narrow range around $\alpha=1$~\cite{Sakai}.

\subsubsection{Integer spins}

An interesting application for integer spins is a possible extension
of Haldane's conjecture to the quantum spin tube. Reconsidering again
the Berry phase term, we examine the possibility of rewriting the sum
of integrals (\ref{eq:Berry3}) into a single one:
\begin{align}
&  i\frac{S}{2} \frac{1-\alpha}{\alpha} \sum_{i = 0}^{P-1} \int_{x_i}^{x_{i+1}} \ud x \int \ud \tau \; \partial_x \mathbf{e}_3 \cdot ( \partial_\tau \mathbf{e}_3 \times \mathbf{e}_3 ) \nonumber \\
 &\equiv  i\frac{S}{2}  \frac{1-\alpha}{\alpha} \int_{-\infty}^{+\infty} \ud x \int \ud \tau \; \partial_x \mathbf{e}_3 \cdot ( \partial_\tau \mathbf{e}_3 \times \mathbf{e}_3 ).
\label{eq:continuous}
\end{align}
Because of (\ref{eq:supequal}), we saw that the integral on the right
hand side of (\ref{eq:continuous}) must vanish for any smooth
configuration of the field $\R(x,t)$. However, it is possible that the
field $\mathbf{e}_3(x,t)$ is smooth but that $\R(x,t)$ is not (see for
instance fig \ref{fig:instantons}: the vectors $\mathbf{e}_1$ and
$\mathbf{e}_2$ change sharply of direction where the instantons take
place but $\mathbf{e}_3$ remains constant). In this case, writing the
Berry phase as a single integral is allowed and this integral will be
different from zero. However, we must emphasize that the
identification (\ref{eq:continuous}) is not totally complete, since we
elude all the instantons events where $\mathbf{e}_3$ is
discontinuous. However, in the region $\alpha \sim 0.5$ we saw that in
the bare action the stiffness of the $\mathbf{e}_1$ is very small. One can then suppose that in this limit the low
energy configurations with non-trivial topological index are those
where the $\mathbf{e}_3$ is smoothly varying and the necessary
discontinuities are in the $\mathbf{e}_1$ field configuration.
Keeping this picture even for larger values of $\alpha$, from
(\ref{eq:continuous}) one then recognizes a topological term for the
unit vector $\mathbf{e}_3$. But this time, it is multiplied by a
factor $(\alpha-1)/\alpha$. For the $O(3)/O(2)$ NL$\sigma$M, it is
known that such a term would lead to a significant change in the
spectrum of the model if it is an half-integer, in which case the
NL$\sigma$M is gapless. Here, we find $2S$ particular values of
$\alpha$ for which this happen:
\begin{eqnarray}
\alpha_p &=& \frac{S}{S-(p+\frac{1}{2})} \rightarrow S_{BP}^\prime = i\pi(2p+1) \\
& & -S < p + \frac{1}{2} < S, \; p \in \mathbb{Z}, \nonumber
\end{eqnarray}
the last inequalities coming from the condition $\alpha > 0.5$. For $\alpha = \alpha_p$, the Berry phase reduces again to an odd multiple of $\pi$. Finally, the full anisotropic sigma model at these points read:
\begin{eqnarray}
S_{\rm NL\sigma M} &=& \int \ud x \ud \tau \left( \frac{1}{2\tilde{g}_a} \left(\tilde{c}_a(\partial_\tau \mathbf{e}_a)^2+\frac{1}{\tilde{c}_a}(\partial_x \mathbf{e}_a)^2 \right) \right. \\
&+& \left. \kappa (\mathbf{e}_3 \cdot \partial_\tau \mathbf{e}_1)^2 + i\frac{2p+1}{4}\partial_x \mathbf{e}_3 \cdot ( \partial_\tau \mathbf{e}_3 \times \mathbf{e}_3) \right).  \nonumber
\end{eqnarray}

Would the RG analysis of the preceding section have predicted a decoupling of the field $\mathbf{e}_1$, we would have concluded that the last equation represents $2S$ critical field theories, each corresponding to an $SU(2)_1$ Wess Zumino Novikov Witten (WZNW) model, as it happens for a dimerized spin $S$ chain~\cite{Affleck}. However, the RG results suggest the opposite scenario: the coupling
between the $\mathbf{e}_1$ and the $\mathbf{e}_3$ is a relevant
perturbation and any non-zero skyrmion configuration must come with a
fugacity, corresponding to the energy cost of a discontinuity of the
$\mathbf{e}_1$ field configuration. This would exclude the scenario for
a $SU(2)_1$ WZNW criticality. Although the one-loop RG results are well
suited to study the vicinity of the point $\alpha =1$, one can
question their validity in the quasi-collinear regime $\alpha \sim
0.5$. The nature of the transition points
$J \to J+1$ ($J$ is the total spin per triangle, see equation
(\ref{eq:sequence1}) is thus unclear for the case $J$ big (i.e. close
to $\alpha=0.5$) and we are going now to study the transition $J=0$ to
$J=1$ in the case of an $S=1$ tube.

\section{\label{sec: S=1}
The $S=1$ case}

In this section, we will focus on the spin-1 case for which two
quantum phase transitions are expected (close to $\alpha=1/2$ and
$\alpha=2$ respectively when $J_\parallel/J_\perp\ll 1$).

\subsection{\label{subsec: Effectivemodel}
Effective model for the $S=1$ tube}

In the small $J_\parallel$ limit, we can apply simple perturbation
theory in order to obtain an effective model which should be valid
close enough to a critical point. As recalled in Sec.~\ref{sec:
  Modelandlimit}, a single triangle exhibits at low-energy a level
crossing between one singlet and one triplet states both at
$\alpha=1/2$ and $\alpha=2$. If we restrict ourselves to the
neighboring of one level crossing, then we can build an effective
model by keeping only these low-lying degrees of freedom. Because the
Hilbert space of one singlet plus one triplet is equivalent to two
spin 1/2, we prefer to describe the effective model in terms of
effective spin 1/2 variables so that the effective model of the
spin-tube becomes a spin-1/2 2-leg ladder hamiltonian.

By performing first order (in $J_\parallel$) degenerate perturbation,
we end up with a SU(2) spin-1/2 ladder that only contains 2-spin
exchange interactions of the form~:
\begin{eqnarray}\label{Heff.eq}
\hat{H}_{\mbox{eff}} & = &
\sum_i \tilde{J}_\perp \tilde{S}_{i,1} \cdot \tilde{S}_{i,2} + 
 \tilde{J}_\parallel (\tilde{S}_{i,1} \cdot \tilde{S}_{i+1,1} + \tilde{S}_{i+1,2} \cdot \tilde{S}_{i+1,2}) \nonumber \\
 & + & \tilde{J}_d \sum_i (\tilde{S}_{i,1} \cdot \tilde{S}_{i+1,2} + \tilde{S}_{i+1,2} \cdot \tilde{S}_{i,1})
\end{eqnarray}
and the effective exchange are as follows:
\begin{align}
\alpha\simeq 1/2: & \quad \tilde{J}_\perp=2\alpha-1, \quad \; \tilde{J}_\parallel=(\frac{11}{8}+\frac{5}{9})J_\parallel, 
 \notag \\
& \tilde{J}_d = (\frac{11}{8}-\frac{5}{9})J_\parallel
   \sim  0.82 J_\parallel .
\label{eq:Jd1}
\\
\alpha \simeq 2: &  \quad \tilde{J}_\perp=2-\alpha, \quad \;
 \tilde{J}_\parallel=\frac{13}{9}J_\parallel,
\notag \\
& \quad  \tilde{J}_d=\frac{5}{9}J_\parallel
 \sim 0.56 J_\parallel .
\label{eq:Jd2}
\end{align}
This mapping allows for a straightforward explanation of the occurence
of quantum phase transitions.  Varying $\alpha$ is equivalent to
changing the effective rung exchange from strongly positive to
strongly negative, which means that the spin-1/2 ladder is in a
rung-singlet phase on one side and in a Haldane phase on the other
side. Because this spin-1/2 model is simpler and has already been
studied intensively, we can use some results from the literature to
clarify the nature of the phase transition.

From the bosonisation point of view, which is valid when
$\tilde{J}_\parallel$ is the dominant energy scale, Nersesyan and
Tsvelik have argued that there should be a transition when
$\tilde{J}_\perp=2\tilde{J}_d$ with the possibility of deconfined
spinons~\cite{Nersesyan2003}.  A more refined analysis by Starykh and
Balents has shown that marginal interactions modify these conclusions
so that the transition between rung singlet and Haldane phase becomes
either first order or has an intermediate columnar dimer
phase~\cite{Starykh2004}. These estimates $\tilde{J}_\perp=2\tilde{J}_d$
for the quantum phase transition are plotted on the phase
diagram in Fig.~\ref{PhasediagS=1}.

From the numerical point of view, early DMRG simulations~\cite{Wang2000}
were interpreted in favor of a second order (respectively first order)
phase transition for $\tilde{J}_d/\tilde{J}_\parallel$
smaller (respectively larger) than 0.287.
The absence of an intermediate dimerized phase was confirmed
by more recent numerical work~\cite{Hung2006,Kim2008} although
these studies do not agree on the order of the transition: either it is
always first order~\cite{Kim2008} or it could be continuous for small
$\tilde{J}_d/\tilde{J}_\parallel \sim 0.2$~\cite{Hung2006}. 
Given that our effective models have a relatively large ratio
$\tilde{J}_d/\tilde{J}_\parallel$
(respectively close to 0.42 and 0.38
for both critical cases $\alpha \sim 1/2,2$),
all numerical studies agree that the phase transitions are
first order.

\subsection{\label{subsec: DMRG}
DMRG results for the $S=1$ tube}

In order to have an unbiased answer, we have decided to perform numerical simulations of the $S=1$ tube with the powerful DMRG algorithm~\cite{dmrg} for several values of $J_\parallel$. Simulations are done mostly with open boundary conditions (OBC) with system sizes up to $3\times 64$, but also with periodic boundary conditions (PBC) on some cases. Typically, we keep up to 1600 states, which is sufficient to have a discarded weight smaller than $10^{-8}$. 

\subsubsection{\label{subsubsec: stringorder}
String order parameter}

In order to draw a numerical phase diagram and to compare it with the
conjectured one (see Fig.~\ref{PhasediagS=1}), we have computed the
$z$-component of the string order parameter (see
Eq.~(\ref{eq:string-order2})) for several values of $J_\parallel$ and
$\alpha$. In order to extract the bulk value and avoid finite-size
effects due to the edges, we have taken the following definition in
our simulations:
\begin{equation}\label{eq:string_dmrg}
\left\langle \mathcal{O}^z \right\rangle
= \left \langle J^z_{L/4} \exp \left( i \pi \sum_{l = L/4}^{3L/4} J_l^z \right) J^z_{3L/4+1} \right \rangle 
\end{equation}

\begin{figure}
\begin{center}
\includegraphics*[width=8cm]{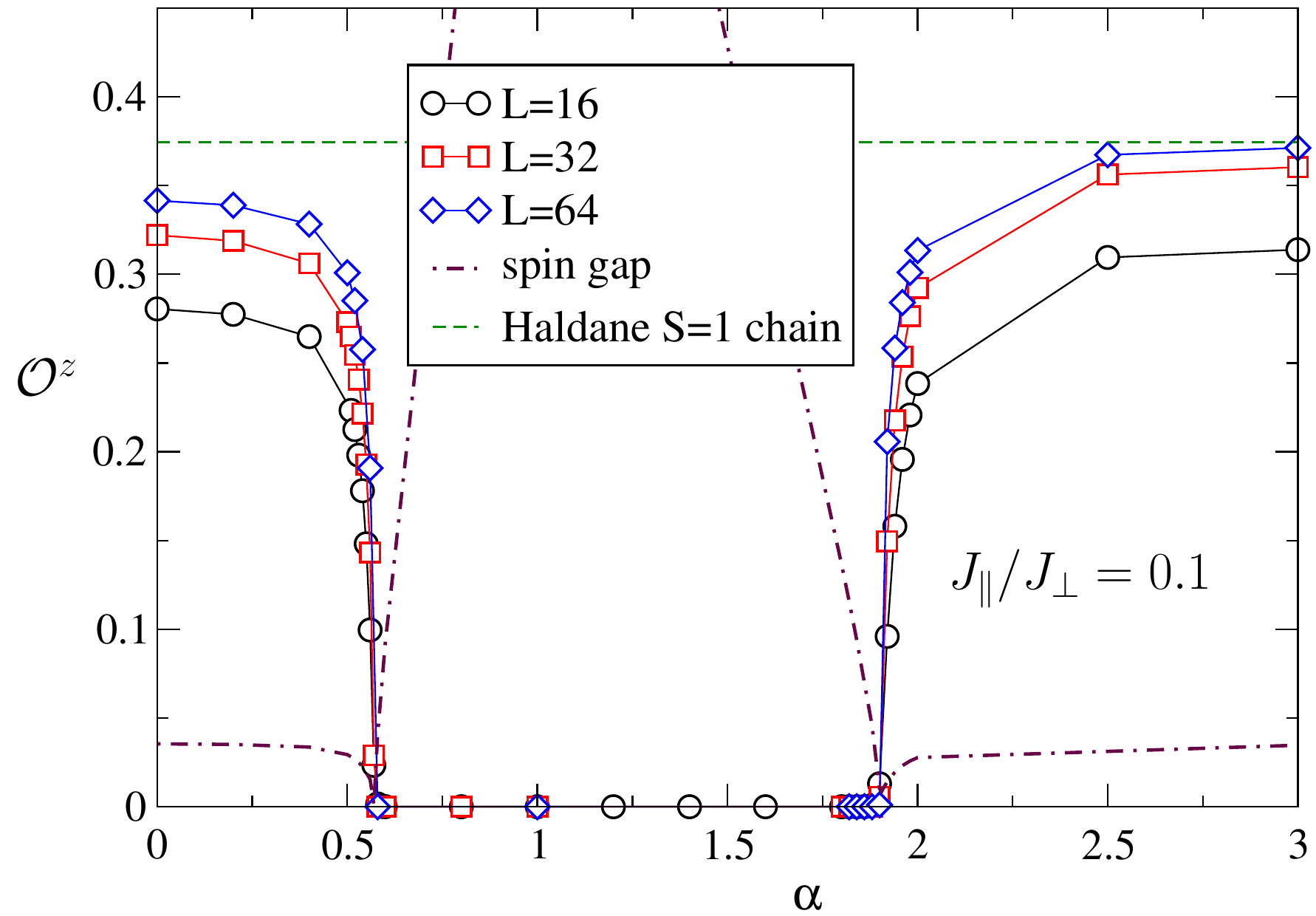}
\caption{(color online) String order parameter along $z$ of the spin-1 tube as a function 
of the anisotropy parameter for $J_\parallel = 0.1 J_\perp$ and various lengths $L$. 
The extrapolated spin gap (see Fig.~\ref{fig:spingap} below) is also shown, as well as the non-local order parameter of the spin-1 Haldane chain.}
\label{fig:stringorder01}
\end{center}
\end{figure}

In Fig.~\ref{fig:stringorder01}, we plot this quantity as a function of
the frustration $\alpha$ for a small $J_\parallel/J_\perp=0.1$.  We
conclude that the string order is finite for $\alpha<0.57$ and
$\alpha>1.92$, and it vanishes elsewhere, i.e. our model does exhibit
quantum phase transitions. 
Therefore,  as a function of $\alpha$, we find successively a
topological phase (with $S_b=1/2$ in the presence of OBC), a
non-topological one, and again a topological phase (with $S_b=1/2$ in
the presence of OBC). 
This is the behaviour expected from the perturbation and the mapping to
an effective spin-1 or spin-0 chain. 
Indeed, for small or large $\alpha$, we can derive an effective spin-1
Haldane model for which the string order parameter is
known~\cite{white93} to be $\sim 0.374$, which is close to our value in
both limits.

\begin{figure}
\begin{center}
\includegraphics*[width=8cm]{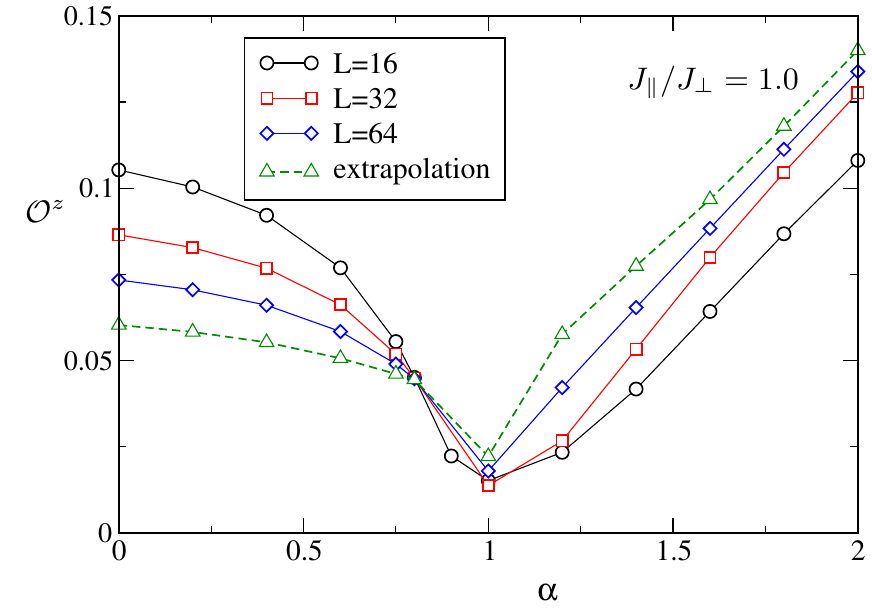}
\caption{(color online) String order parameter along $z$ of the spin-1 tube as a function 
of the anisotropy parameter for $J_\parallel =  J_\perp$ and various lengths $L$. A linear extrapolation with the largest sizes is also plotted and is finite for all $\alpha$.}
\label{fig:stringorder10}
\end{center}
\end{figure}

In order to complete our phase diagram in Fig.~\ref{PhasediagS=1}, we also compute the string order parameter for larger $J_\parallel/J_\perp$ where perturbation is no more valid. Data are shown in Fig.~\ref{fig:stringorder10} and have a quite different behaviour: now, the string order parameter is finite for all $\alpha$, i.e. we have no phase transitions along this line.
 
By computing $\langle \mathcal{O}^z \rangle$
for various $J_\parallel$ and $\alpha$, we estimate that the tip
of the $J=0$ lobe occurs for $J_\parallel/J_\perp=0.67$ and $\alpha=1$. 

\begin{figure}
\begin{center}
\includegraphics*[width=8cm]{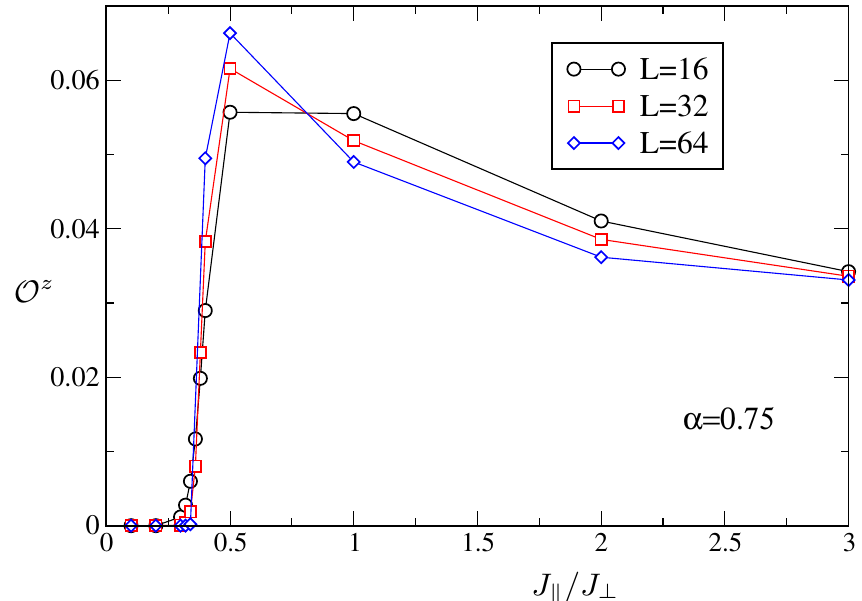}
\caption{(color online) String order parameter along $z$ of the spin-1 tube as a function 
of  $J_\parallel/J_\perp$ for a fixed anisotropy parameter $\alpha=0.75$ and various sizes.}
\label{fig:stringorder075}
\end{center}
\end{figure}

Now, we present data for a vertical cut in the phase diagram of
Fig.~\ref{PhasediagS=1} by fixing $\alpha=0.75$. By varying
$J_\parallel/J_\perp$, the string order plotted in
Fig.~\ref{fig:stringorder075} vanishes for $J_\parallel/J_\perp\leq
0.34$ and is finite beyond.
In Fig.~\ref{fig:stringorder075},
the string order parameter $\langle \mathcal{O}^z \rangle$
remains finite up to $J_\parallel/J_\perp \sim 3$.
However, as we have discussed in Sec.~\ref{sec:hiddenZZ_weak},
$\langle \mathcal{O}^z \rangle$ vanishes in the
weak coupling limit.

\begin{figure}
\begin{center}
\includegraphics*[width=8cm]{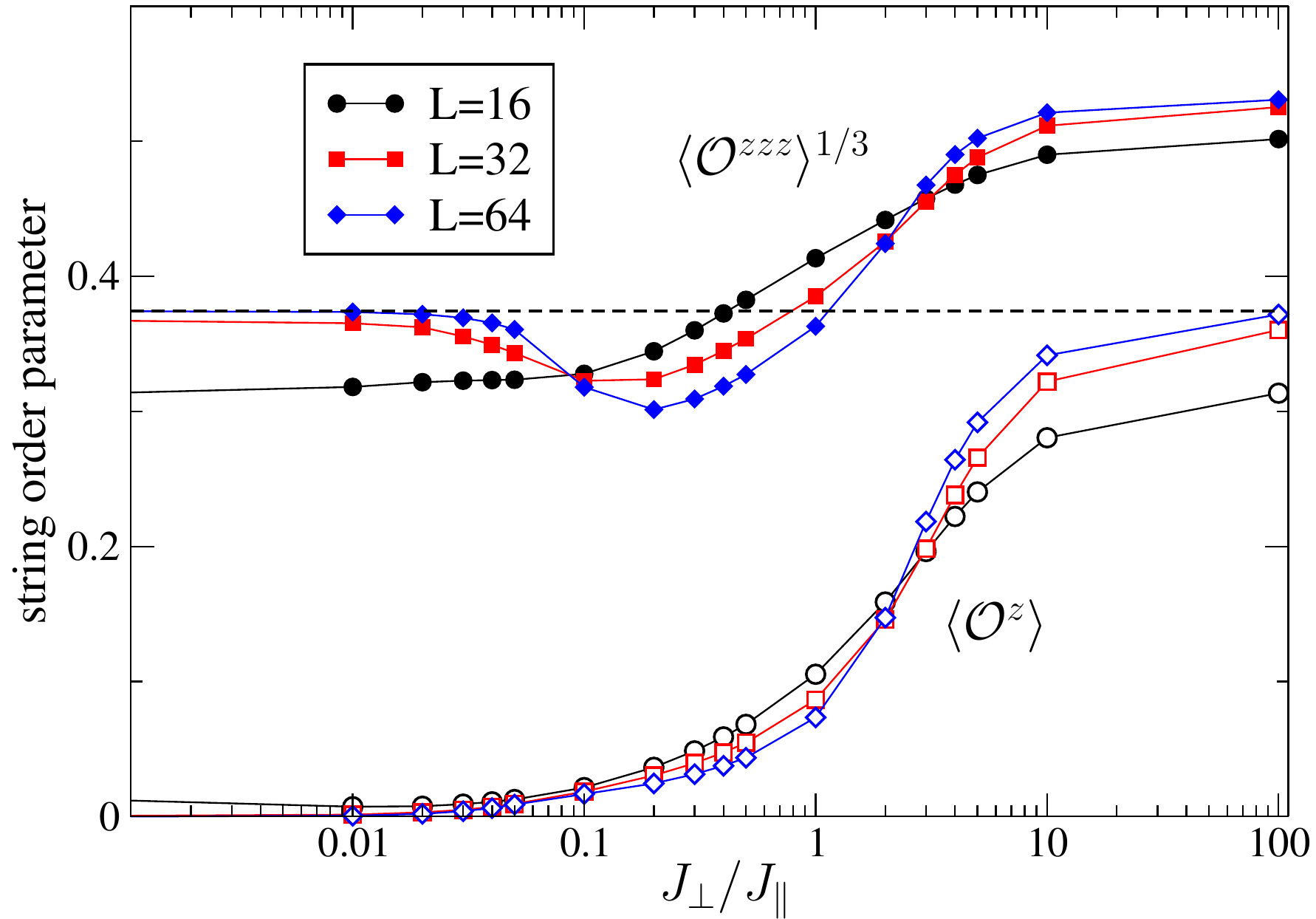}
\caption{(color online) Various string order parameters of the spin-1 ladder ($\alpha=0$)
 as a function of  $J_\perp/J_\parallel$ for various sizes. Usual (respectively product) 
string order along $z$ is shown with open (respectively filled) symbols. The dashed line indicates the known value for the spin-1 Haldane chain~\cite{white93}.} 
\label{fig:stringorderv2}
\end{center}
\end{figure}

In order to ascertain that there is no other phase transition
when going to the decoupled chain limit, we have computed the product
string order parameter of Eq.~(\ref{eq:product-string}) as well as the
usual one for the non-frustrated case $\alpha=0$. Data are shown in
Fig.~\ref{fig:stringorderv2} for both order parameters. When the chains
are almost decoupled, the product string order parameter is close to the
product of the standard string order parameter for the independent
chains. On the opposite side, when the dominant coupling is $J_\perp$,
the perturbative argument that we have given above (see
Sec.~\ref{subsec: Effectivemodel}) indicates that the 3-leg ladder
behaves effectively as a spin-1 chain for which the string order
parameter $\langle \mathcal{O}^z \rangle$ reduces to the usual one; this
is indeed what is found numerically on large system size.

\subsubsection{\label{subsubsec: Spectralgap}
Spectral gap} 

\begin{figure}
\begin{center}
\includegraphics*[width=8cm]{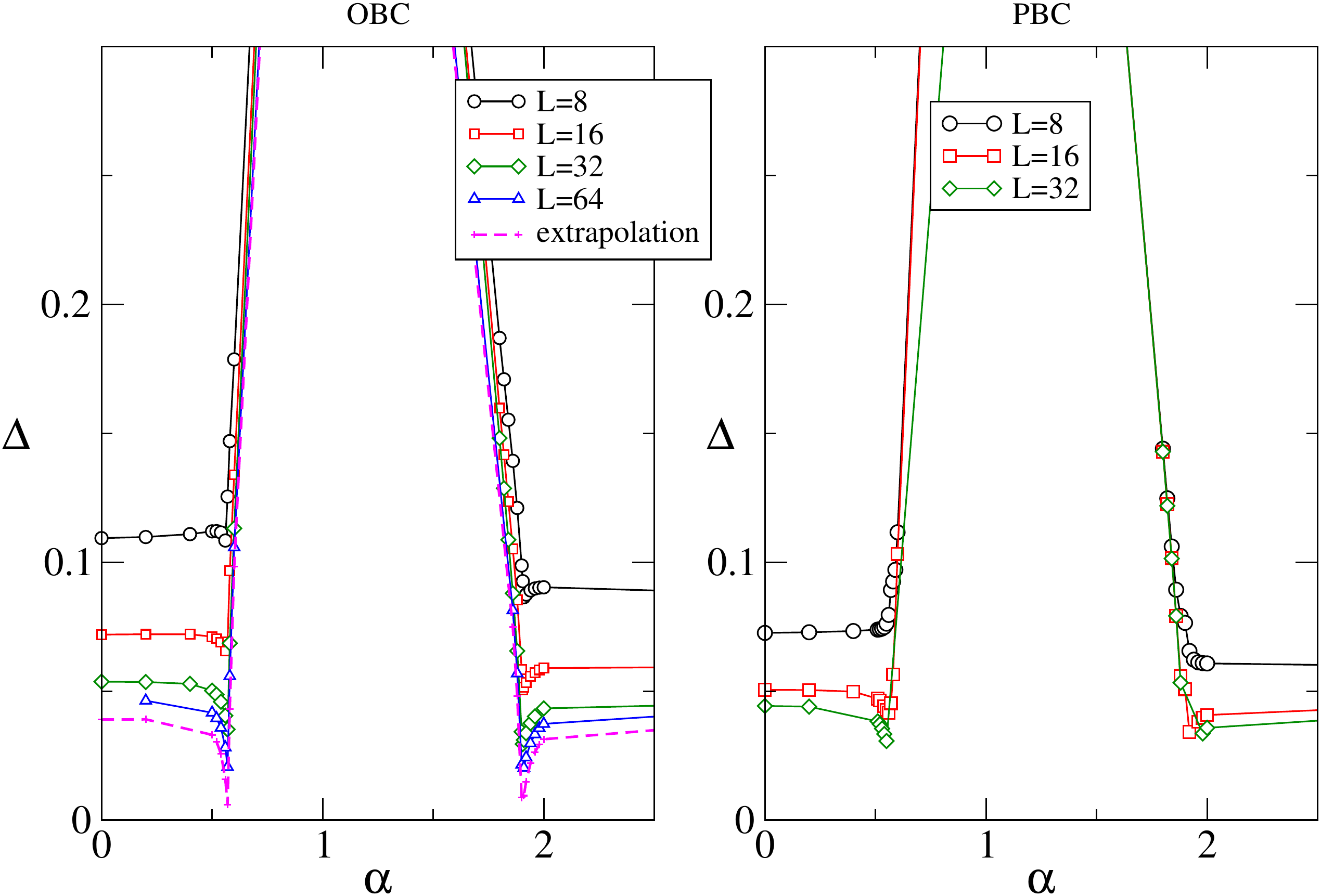}
\caption{(color online) Spin gap of the spin-1 tube as a function of the anisotropy parameter. $J_\parallel = 0.1 J_\perp$ and both
 open and periodic boundary conditions (OBC/PBC) are considered.}
\label{fig:spingap}
\end{center}
\end{figure}

We have also calculated the excitation gap by DMRG.
In order to estimate the bulk gap,
the gap is extracted differently from the finite-size spectrum 
depending on the boundary conditions.
For open boundary conditions (OBC), in a Haldane-like
phase, the real gap should be calculated between the $S = 1$ sector
and the $S = 2$ sector since the sectors $S = 0$ and $S = 1$ are
already degenerate because of the edge states. On the contrary, for a
singlet-like state, the gap is defined between the $S = 0$ sector and
the $S = 1$ sector. For periodic boundary conditions, there are no
edge states and thus, the gap is uniquely defined to be between $S =
0$ and $S = 1$. The evolution of the gap is presented on Figure
\ref{fig:spingap} (left for OBC and right for PBC). For OBC on
largest system sizes and fixed $J_\parallel/J_\perp=0.1$, the DMRG
indicates that the gap between $S = 0$
and $S = 1$ is almost zero for $\alpha < 0.57$ or $\alpha>1.92$ but is finite in between (data not shown). 
The gap between $S = 1$ and $S = 2$ is plotted on Fig.~\ref{fig:spingap} for $J_\parallel=0.1J_\perp$ and 
exhibits  a striking
difference in three regions. 
For $\alpha < 0.57 $ or $\alpha>1.92$, the gap is finite and almost constant with $\alpha$. In contrast, in the intermediate $\alpha$ region, the gap increases almost linearly away from these critical points. 
Extrapolation
of the data for large system sizes seems to go in favor of a finite gap
everywhere (see Fig.~\ref{fig:spingap}), but note that the extrapolated
gap at the critical points is extremely small. 
Both critical points are identical to the values we had found with the
string order parameter.
We observe that the gap is roughly constant in the
$S_b=1/2$ phase, except in the vicinity of the transition point. 
Note that this is in accordance with the
qualitative picture of the strong coupling limit. For small or very large $\alpha$, the effective model is a spin-1 chain with an effective spin exchange of order $J_\parallel$ (but independent of $\alpha$); therefore, the spin gap essentially depends on the value of
$J_\parallel$ but is independent of the anisotropy parameter $\alpha$. 

\begin{figure}
\begin{center}
\includegraphics*[width=8cm]{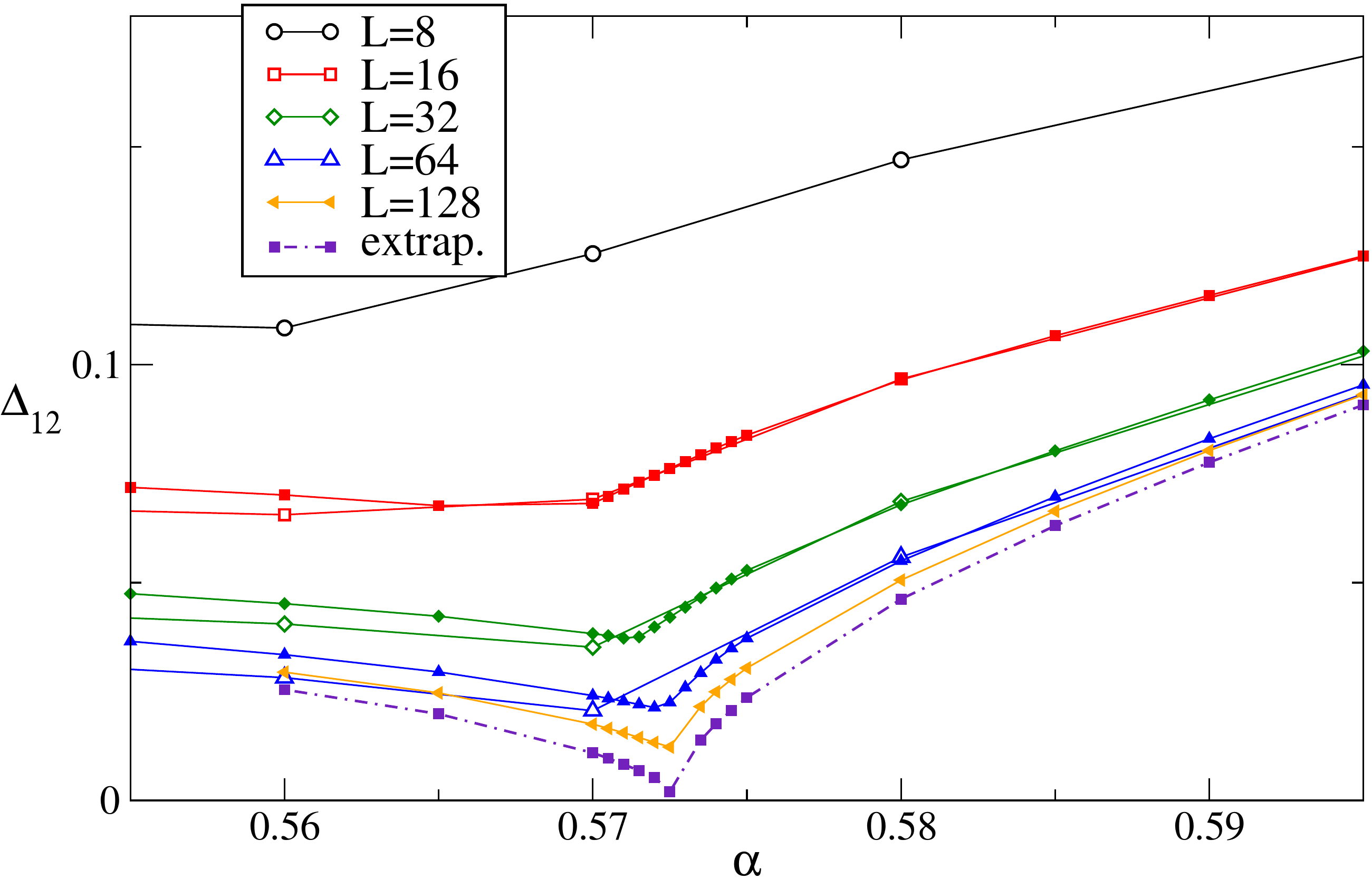}
\caption{(color online) Spin gap of the spin-1 tube as a function of the anisotropy parameter (open symbols) or of the effective
2-leg spin-1/2 ladder (filled symbols). $J_\parallel = 0.1 J_\perp$ and  open  boundary conditions  are used. The infinite-size extrapolation was performed with the two largest ladder clusters.}
\label{fig2_dmrg}
\end{center}
\end{figure}

In order to make connection with the perturbation theory, we have also
performed simulations of the effective spin-1/2 ladder (see
Eq.~(\ref{Heff.eq})) corresponding to our parameters choice. Due to the
Hilbert space reduction, we are able to simulate larger clusters. As
can be seen on Fig.~\ref{fig2_dmrg}, we obtain a very good agreement
between both sets of data since we are indeed considering a small
$J_\parallel$ case where perturbation is expected to be accurate. We
have plotted an infinite-size extrapolation by using the two biggest
ladders ($L=64$ and $L=128$) which confirm that the spin gap has a
large drop around $\alpha=0.57$. However, at the critical point, the
spin gap does not vanish and our data suggest an extremely small but
finite value. This result would indicate a first-order phase
transition, in agreement with other numerical studies on the ladder
systems~\cite{Wang2000,Hung2006,Kim2008}.

\subsubsection{\label{subsubsec: vonNeuman}
Von Neumann entropy}

\begin{figure}
\begin{center}
\includegraphics*[width=7cm]{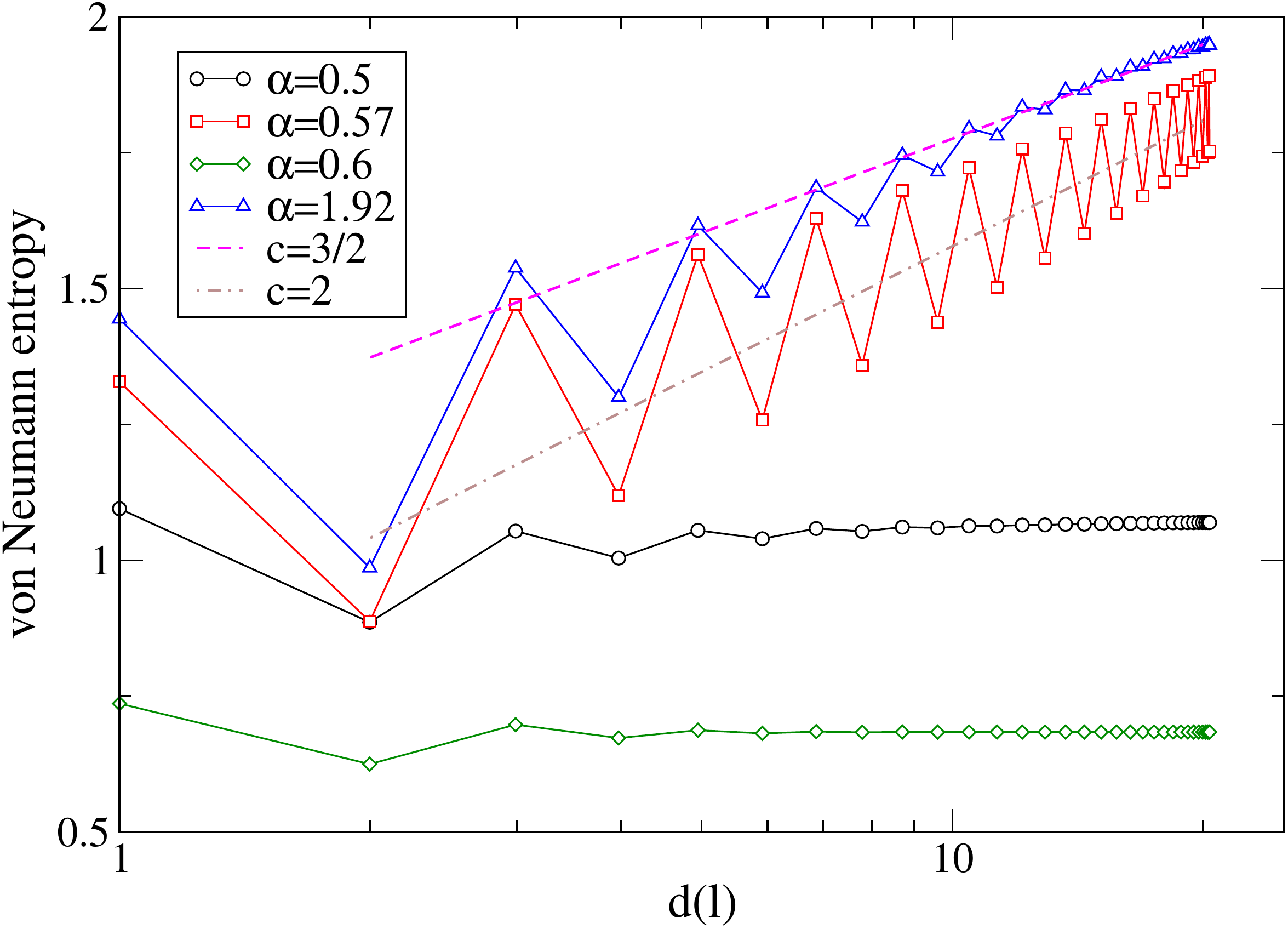}
\caption{(color online) Plot of the Von Neumann entropy as a function of $d(\ell)$ for different values of the anisotropy parameter and fixed  $J_\parallel=0.1J_\perp$.}
\label{fig:spinentropy}
\end{center}
\end{figure}

In this part, we fix $J_\parallel/J_\perp=0.1$ for which we have found
two quantum phase transitions for 
$\alpha=0.57$ and $\alpha=1.92$, both with an extremely small but finite
spin gap at the transition. 
Since these critical points seem to be very close to tricritical points
(the correlation lengths are very large at the transitions), 
another quantity of interest here is the von
Neumann entropy of a finite segment of the chain. It is defined by:
\begin{equation}
S_{vN}(\ell) = -\Tr (\rho_\ell \ln \rho_\ell)
\end{equation}
where $\rho_\ell = \Tr_\ell \rho$ is the reduced density matrix
associated to a block of $\ell$ spins. This quantity is known to
behave fundamentally differently for critical and non-critical
systems~\cite{Vidal,Cardy}. It saturates at a finite value when the
system is non critical while it increases logarithmically for critical
systems. The analytic expression of $S_\ell$ is given by:
\begin{equation}
S_{vN}(\ell) = \frac{c}{6} \ln \left[\frac{2L}{\pi} \sin \left(\frac{\pi \ell}{L} \right) \right] +g
 \end{equation}
 where $c$ is the central charge, and $g$ is a constant. The Von
 Neumann entropy is represented on Figure \ref{fig:spinentropy}  as a 
 function of the conformal distance $d(\ell) = \sin(\pi \ell /L)$. For $\alpha = 0.5$ and
 $\alpha = 0.6$, the entropy converges to a finite value (the same
 results holds for all other values far enough from both critical
 points). On the contrary, for $\alpha = 0.57$ and $\alpha=1.92$,
  it does not saturate but grows with the system size, which should be in favor of 
 a gapless character of the two critical points. 
 The entropy displays also a large periodic oscillation. Such an oscillation has
 already been observed in other critical spin chains~\cite{Laflo2006} and may be
 related to the existence of soft modes at $k = 0$ and $k = \pi$ in
 the problem~\cite{Legeza}. Thus, the decay in correlation function
 would not be simply algebraic at the critical point but the decaying
 function would be multiplied by an oscillatory factor. Because of the
 oscillations, it is hard to distinguish what is the best fit between
 $c = \frac{3}{2}$ and $ c = 2$. 

However, from our DMRG data on the gap, as well as the mapping to an effective ladder, we conclude to a 
very weak first-order transition, especially in the strong coupling
limit $J_\parallel \ll J_\perp$. This last scenario is supported by the
bosonisation studies and the DMRG simulation of the effective
Hamiltonian (\ref{Heff.eq}). Still, von Neumann entropy exhibits a critical behaviour with $c=3/2$ or $c=2$ at the critical 
points, which is valid on rather large length scales. Therefore, we believe that this spin tube can be tuned very 
close to a tricritical point separating a first-order transition line from a possible continuous transition line nearby.
 The gapless transition at the critical
points could then be correlated with the presence of a non trivial
topological term in the NL$\sigma$M. Note that there are no such
critical points in the phase diagram of the two-leg ladder with $S =
1$~\cite{Todo,Allen2000}, and that there is no topological term in the
corresponding NL$\sigma$M as well~\cite{Senechal2}. Of course, the
link between the NL$\sigma$M and the critical point of the DMRG data
still needs to be clarified with further investigations, both analytically and numerically.

\section{\label{sec: S=2}
The $S=2$ case}

In this section, we consider the spin-2 case for which four quantum phase transitions are expected as a function of the frustration $\alpha$ for a fixed $J_\parallel/J_\perp\ll 1$. 

\subsection{\label{subsec: Effectivemodel2}
Effective model for the $S=2$ tube}

In principle, one can apply simple perturbation theory in order to obtain an effective model
valid close enough to any of the critical points for small $J_\parallel/J_\perp$. Here, we have two
types of critical points: (i) close to $\alpha=1/2$ or $\alpha=3$, and as recalled in Sec.~\ref{sec: Modelandlimit}, a single triangle will have a quintet and a triplet as low-energy states; 
(ii) close to $\alpha=2/3$ or $\alpha=3/2$, low-energy states consist in one triplet and one singlet. 

Although one can derive both kinds of effective models, case (i) does not allow to make analytical progress. On the contrary, for the second case, the effective model turns out to be formally
the same as for the spin-1 tube (see part~\ref{subsec: Effectivemodel}), i.e. first-order degenerate perturbation results can be mapped onto a SU(2) spin-1/2 ladder that only contains 2-spin
exchange interactions of the form given in Eq.~(\ref{Heff.eq}). The effective exchanges
 are given as:
\begin{eqnarray}
\alpha& \simeq &2/3: \quad \tilde{J}_\perp=3\alpha-2, \quad \; \tilde{J}_\parallel=\frac{23}{5} J_\parallel, 
 \nonumber \\
& & \tilde{J}_d=\frac{7}{5}J_\parallel . \\
\alpha&\simeq &3/2: \quad \tilde{J}_\perp=3-2\alpha, \quad \; \tilde{J}_\parallel=\frac{151}{40}J_\parallel, \quad  \tilde{J}_d=\frac{39}{40}J_\parallel. \nonumber
\end{eqnarray}
As explained in details in part~\ref{subsec: Effectivemodel}, such a
mapping explains the occurence of a quantum phase transition when
$\tilde{J}_\perp \simeq 2\tilde{J}_d$, as well as giving 
insight on the order of the transition. 

Both quantum critical lines  are plotted in
Fig.~\ref{fig:PhasediagS=2}. Moreover, 
for both cases we are in a regime where
$\tilde{J}_d/\tilde{J}_x \sim 0.25-0.30$, for which 
there is no consensus yet on the order of the
transition~\cite{Hung2006,Kim2008}. Still, if the transitions are
first-order, the gap at the transition should be very small, which means
that for most practical purpose, the system will appear critical on
length scales smaller than the (large) correlation length.

\subsection{\label{subsec: DMRG2}
DMRG results for the spin $S=2$ tube}

Simulations are done mostly with open boundary conditions (OBC) with
system sizes up to $3\times 64$, but also with periodic boundary
conditions (PBC) on some cases. Typically, we keep up to 1600 states,
which is sufficient to have a discarded weight smaller than $10^{-6}$.

We have computed several spectral gaps with various boundary
conditions and various total spin sector, in order to avoid edge
effects, but finite-size effects are rather large and spin gap values
quite small so that no definite answer on the phase diagram can be
obtained this way.

\begin{figure}
\begin{center}
\includegraphics*[width=7cm]{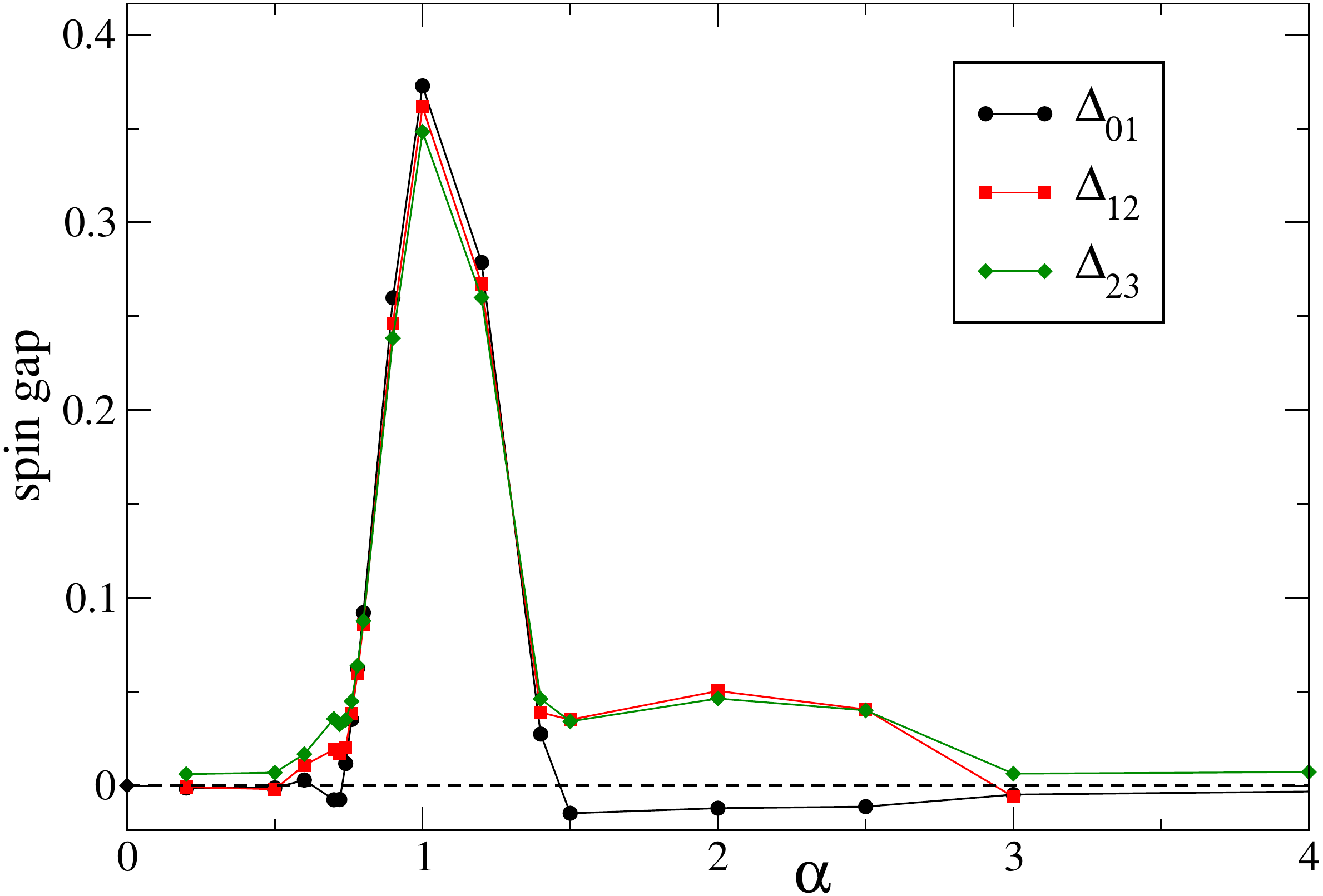}
\caption{(color online) Various spin gaps $\Delta_{ab}$ for the spin-2 tube as a function of $\alpha$ 
for $J_\parallel/J_\perp=0.1$. Data are extrapolated to the thermodynamic limit from simulations with systems of length $L=8$, $16$, and $32$.}
\label{fig:spingap_S2}
\end{center}
\end{figure}

A clearer signal is given by the string order parameter (see
Eq.~\ref{eq:string_dmrg}) which can distinguish between odd-$J$ and
even-$J$ Haldane phases. Unfortunately, this quantity alone cannot
distinguish between $J=0$ and $J=2$ phase but we can rely on the
presence/absence of edge states to distinguish these phases.  In
Fig.~\ref{fig:spingap_S2}, we plot various spin gaps as a function of
the frustration $\alpha$. Obviously, if edge states carry a spin
$S_b$, then the bulk spin gap is obtained between lowest levels of
total spin $(2S_b)$ and $(2S_b+1)$, $\Delta_{2S_b, 2S_b+1}$, i.e. once
edge states have been polarized. In the thermodynamic limit, the two
$S_b$ edge states form $(2S_b+1)^2$ degenerate states. Our data are
obtained on finite lengths up to $L=32$ and we perform a $1/L$ linear
extrapolation to get an estimate of the thermodynamic limit spin
gaps. For a small $J_\parallel/J_\perp=0.1$, our data shown in
Fig.~\ref{fig:spingap_S2} indicate successively regions with $S_b=1$,
$1/2$, $0$ and then in reverse order for increasing $\alpha$, as had
been conjectured initially in section~\ref{subsec: Edgestates}.

\begin{figure}
\begin{center}
\includegraphics*[width=8cm]{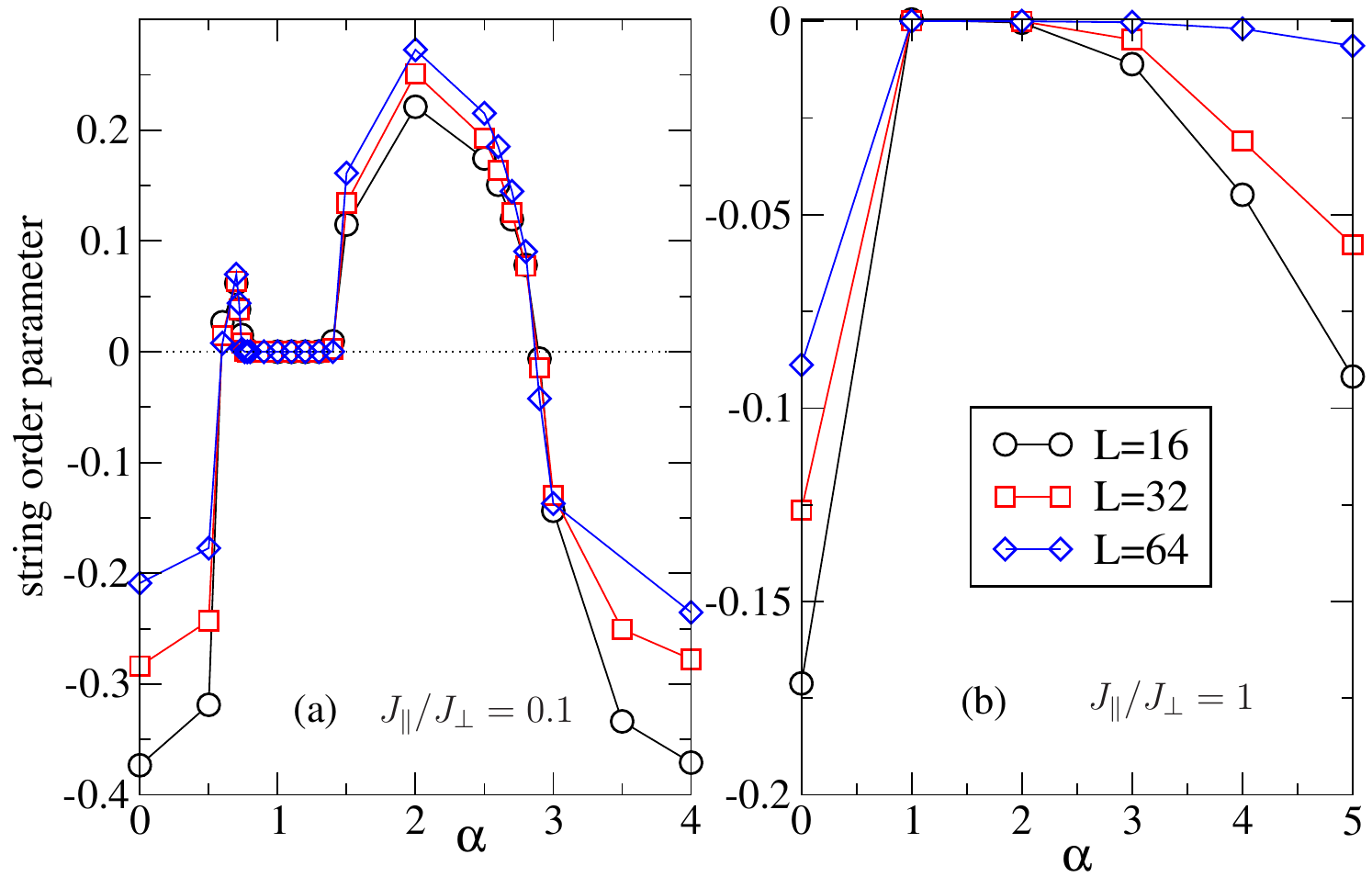}
\caption{(color online) String order for the spin-2 tube as a function of $\alpha$ 
for various lengths $L$ with (a) $J_\parallel/J_\perp=0.1$ or (b) $J_\parallel/J_\perp=1$.}
\label{fig:stringS2}
\end{center}
\end{figure}

\begin{figure}
\begin{center}
\includegraphics*[width=8cm]{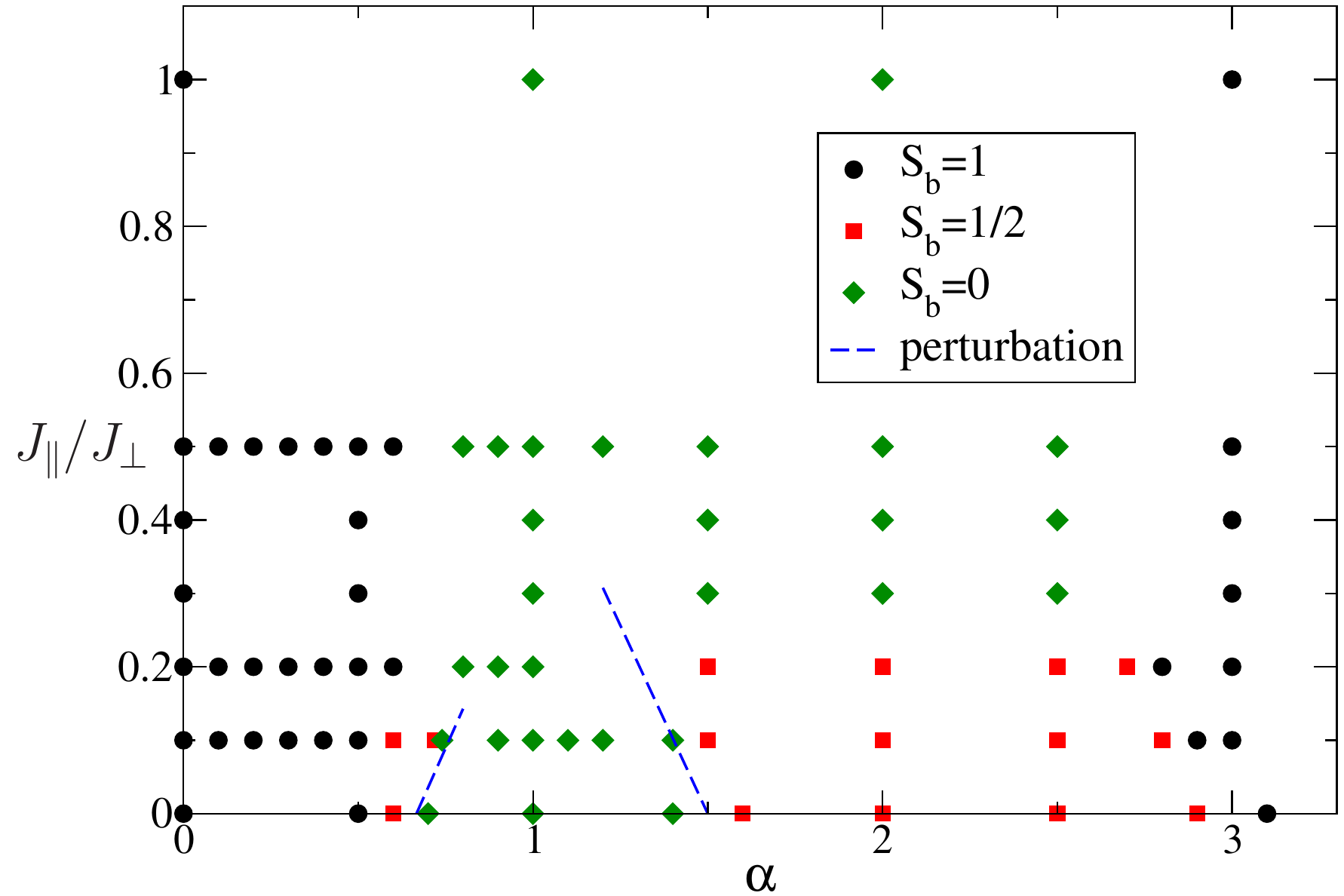}
\caption{(color online) Numerical phase diagram for the spin-2 tube obtained from the spin gap and the string order parameter. }
\label{fig:phasediagS2}
\end{center}
\end{figure}

In order to get more insight in this phase diagram, we have computed
the string order parameter (see Eq.~\ref{eq:string_dmrg}) for several
exchange couplings. Data are shown in Fig.~\ref{fig:stringS2} and
confirm the existence of even/odd Haldane S phases: a spin 0 phase has
a vanishing order parameter already on finite systems (see $\alpha$
close to 1); a spin 1 phase has a finite and positive string order
parameter; a spin 2 phase should have a zero string order parameter
but the finite-size scaling is rather slow, as is already known for
the spin-2 Haldane chain for instance~\cite{Yamamoto1997}.  Note that
the string order data are perfectly consistent with the edge states
picture drawn from the spin gap data. This way, one can draw a
quantitative phase diagram for the spin 2 tube in
Fig.~\ref{fig:phasediagS2}, which confirms the qualitative picture
conjectured in Fig.~\ref{fig:PhasediagS=2}.

\section{\label{sec: Conclusion}
Discussion and Conclusion}

In this paper we have shown that a simple model such as a three-leg
quantum spin ladder
can give rise to a very rich phase diagram. 
As it is now ubiquitous in quantum spin chains, integer and half integer
spin must be treated separately. 
For half-integer spins, the Berry phase analysis of
Section~\ref{sec:Berry} 
points towards a quantum phase transition between
a gapped
spectrum and a degenerate ground state for $\alpha$ close to $1$ and a
gapless regime on each side of this phase, as it has indeed being
observed for the $S=1/2$ case~\cite{Sakai}. The semiclassical picture of
this scenario is a phase transition separating a gapped isotropic
coplanar phase and a pseudo-collinear gapless regime. Not surprisingly
the difference in behavior at large scales is dictated by the Berry
phase terms present in the action.

For integer spins the situation is even more interesting. If we
consider the strong coupling regime $J_{\perp} \gg J_{\parallel}$ it
makes no doubt that $2S$ quantum phase transitions are expected
for a spin $S$ tube when varying the anisotropy parameter
$\alpha$.
These phase transitions separate gapped phases, and this
scenario is reminiscent of what happens in dimerized spin chains when
varying the dimerization parameter (see for example I. Affleck's
lectures~\cite{Affleck}).

These phase transitions can be understood in terms
of spontaneous breaking of the hidden \ZZ\ symmetry. 
The broken hidden \ZZ\ symmetry can be detected by
a string order parameter or by edge states with half-integer spin.
While a simple generalization of the
string order parameter to the tube vanishes in the
weak rung coupling limit, an alternative string order parameter
remains finite in the same limit for odd spin $S$.
Additionally, in some regions of the phase diagram, there exist ``phases''
with integer-spin edge states.
Although they appear to be non-trivial phases,
they can be adiabatically connected to a trivial phase
with no edge state.
This is consistent with an unbroken hidden \ZZ\ symmetry
in these ``phases''.

More insight into the phase diagram can be obtained with
the recent discussion concerning the characterization of
the Haldane phase~\cite{GuWen,Pollmann09a,Pollmann09b}.
The Haldane phase can remain a distinct
phase separated from a trivial phase by a quantum
phase transition, even when the hidden \ZZ\ symmetry
is not well-defined and the string order parameter
is not useful.
It turns out that the Haldane phase is a
topological phase protected by any one of the following
three symmetries: 1) global D$_2$ ($=$ \ZZ ) symmetry
of $\pi$-rotation about $x,y$ and $z$ axes,
2) time-reversal symmetry (for $\vec{S}_j \rightarrow - \vec{S}_j$),
and 3) lattice inversion symmetry about a bond center (link parity).
The hidden \ZZ\ symmetry is well-defined only with the symmetry 1)
above.
Most generally, the Haldane phase is characterized by
an exact double degeneracy of the entanglement spectrum.

In this paper, we limited our discussion to the tube
with SU(2) symmetry of spin rotation and all the symmetries
1)--3) listed above.
Within this limitation,
the hidden \ZZ\ symmetry can be used to characterize
the nontrivial phases, which have edge states with a half-integer
spin $S_b$.
On the other hand, we expect that the phases with
the broken hidden \ZZ\ symmetry correspond to the
generalized Haldane phase with an exact double
degeneracy of the entanglement spectrum.
It is protected by either of the symmetries 2) or 3),
even when the symmetry 1) is explicitly broken
and the hidden \ZZ\ symmetry is ill-defined.
In particular, as long as the lattice inversion symmetry
abount a bond center is preserved,
the generalized Haldane phase is protected as
a distinct topological phase.
This protection may be roughly understood by the intrinsic odd
parity with respect to the lattice inversion
associated to an odd number of valence bonds
between the neighboring rungs~\cite{Pollmann09a}.

The above general analysis implies the existence of a quantum
phase transition between a ``topological'' phase
and a ``trivial'' phase.
However, it does not tell the order of the transition
or its universality class.

For dimerized chains, the NL$\sigma$M
approach shows that the critical points correspond to an effective
half-integer chain which is described by an $SU(2)_1$ WZNW model. Our
analysis of the NL$\sigma$M  in the triangular spin tube has shown that the situation is different here indicating
that we must expect phase transition of a different kind. Arises then
the question of whether these transition points are expected
to be first order or more interesting critical theories as for example
higher levels $SU(2)$ WZNW models.

The case of the $S=1$ ladder has proven to be a very interesting and
instructive example. The low energy behavior of this system can be
shown to be equivalent to a two-leg spin $1/2$ ladder. This ladder
system has two obvious extreme regimes corresponding to a collection
of singlet states (strong antiferromagnetic couplings between the chains) and a
Haldane phase of an effective spin $1$ chain (strong ferromagnetic coupling
between the chains). The most recent bosonization analysis~\cite{Starykh2004} indicates that the transition between these two
regimes can be either first order, or a couple of (gapless) lines
surrounding a dimerized phase.  These gapless lines have central
charges $c=1/2$ and $c=3/2$, this last one corresponding to a
$SU(2)_2$ WZNW model. We have performed extensive DMRG computations on
this system. The analysis of the spectral gap and the von Neumann
entropy tend to indicate a weak first order transition for our system,
but in any case the close proximity to the tricritical point. This
allows us to speculate that by introducing further microscopic
parameters, as for example second neighbors interactions, the
transition can be made second order, but this issue is beyond the
scope of the present work. This result is also encouraging for
analyzing the nature of the transition for higher spins with both 
numerical and novel analytical techniques.

One important point is that many of the results obtained here
generalize to ladders with a higher odd number of legs displayed with 
periodic boundary conditions (so in a frustrating manner). Of course
frustration
becomes weaker as one increase the number of legs. In this sense the
3-leg ladder 
is a representative of a family of quasi one-dimensional systems where
frustration
plays a crucial role in the emergence of an interesting physics. 

\begin{acknowledgments}
We would like to thank F. Alet, J. Almeida, P. Azaria, E. Berg,
F. Dahmani, K. Damle, D. Mouhanna, F. Pollmann, G. Sierra, and A.~M. Turner
for enlightening discussions.
S.~C. thanks Calmip (Toulouse) for computing time.
M.~O. is supported in part by JSPS Grant-in-Aid for Scientific Research
(KAKENHI) No. 18540341. 
\end{acknowledgments}


\begin{thebibliography}{99}
\bibitem{note-velocity}
	We have compared the velocities $c_\alpha =
	\sqrt(\rho_\alpha / \chi_\alpha)$ deduced from the NL$\sigma$M
	and the one obtained from a spin-wave expansion at leading
	order. It appears that the values of $c_1$ and $c_2$ slightly
	differs in the two approaches away from $\alpha = 1$. This might
	be due to the difference between Eq.~\protect\eqref{eq:ansatz} and
	Eq.~\protect\eqref{eq:ansatz2}.
	However, we find that the result of  the
	integration of the RG equations is not altered by these
	variations.
\end{thebibliography}
\end{document}